\documentclass[twocolumn]{aastex631}

\usepackage{amsmath}
\usepackage{empheq}
\usepackage{mathrsfs}
\usepackage{textcomp}
\usepackage{enumitem}   
\usepackage{gensymb}
\usepackage{hyperref}
\usepackage{graphicx}
\usepackage[caption=false]{subfig}
\usepackage{multirow}
\usepackage{longtable}
\usepackage{ulem}
\usepackage[flushleft]{threeparttable}
\usepackage{booktabs} 
\usepackage{CJK}
\hypersetup{colorlinks, linkcolor={blue}, citecolor={blue}, urlcolor={blue}} 

\definecolor{DarkOrange}{RGB}{204, 85, 0}
\definecolor{LincolnGreen}{RGB}{17, 102, 0}
\def\ion#1#2{#1$\;${\footnotesize\rm{#2}}\relax}

\usepackage{xspace}

\newcommand\swift{\textit{Swift}\xspace}

\newcommand\srg{\textit{SRG}\xspace}

\def \caltech {{Division of Physics, Mathematics and Astronomy, 
California Institute of Technology, Pasadena, CA 91125, USA}}

\def \coo {{Caltech Optical Observatories, California Institute of Technology, Pasadena, CA 91125, USA}}

\def \ljmu {{Astrophysics Research Institute, Liverpool John Moores University, IC2, Liverpool Science Park, 146 Brownlow Hill, Liverpool L3 5RF, UK}}

\begin{document}
\pagenumbering{arabic}

\title{
A Luminous Red Optical Flare and Hard X-ray Emission \\ in the Tidal Disruption Event AT\,2024kmq 
}


\author[0000-0002-9017-3567]{Anna Y. Q.~Ho}
\affiliation{Department of Astronomy, Cornell University, Ithaca, NY 14853, USA}

\author[0000-0001-6747-8509]{Yuhan Yao}
\affiliation{Miller Institute for Basic Research in Science, 468 Donner Lab, Berkeley, CA 94720, USA}
\affiliation{Department of Astronomy, University of California, Berkeley, 501 Campbell Hall, Berkeley, CA, 94720, USA}

\author[0000-0002-3809-0000]{Tatsuya Matsumoto}
\affiliation{Department of Astronomy, Kyoto University, Kitashirakawa-Oiwake-cho, Sakyo-ku, Kyoto, 606-8502, Japan}
\affiliation{Hakubi Center, Kyoto University, Yoshida-honmachi, Sakyo-ku, Kyoto, 606-8501, Japan}

\author[0000-0001-9915-8147]{Genevieve Schroeder}
\affiliation{Department of Astronomy, Cornell University, Ithaca, NY 14853, USA}

\author[0000-0003-3765-6401]{Eric Coughlin}
\affiliation{Department of Physics, Syracuse University, Syracuse, NY 13210, USA}

\author[0000-0001-8472-1996]{Daniel A.~Perley}
\affiliation{\ljmu} 


\author[0000-0002-8977-1498]{Igor Andreoni}
\affiliation{University of North Carolina at Chapel Hill, 120 E. Cameron Ave., Chapel Hill, NC 27514, USA}

\author[0000-0001-8018-5348]{Eric C. Bellm}
\affiliation{DIRAC Institute, Department of Astronomy, University of Washington, 3910 15th Avenue NE, Seattle, WA 98195, USA}

\author[0000-0001-9152-6224]{Tracy X. Chen}
\affiliation{IPAC, California Institute of Technology, 1200 E. California
             Blvd, Pasadena, CA 91125, USA}

\author[0000-0002-7706-5668]{Ryan Chornock}
\affiliation{Department of Astronomy, University of California, Berkeley, 501 Campbell Hall, Berkeley, CA, 94720, USA}

\author[0000-0003-1858-561X]{Sofia Covarrubias}
\affiliation{\caltech} 

\author[0000-0001-8372-997X]{Kaustav Das}
\affiliation{\caltech} 

\author[0000-0002-4223-103X]{Christoffer Fremling}
\affiliation{\coo} 
\affiliation{\caltech} 

\author{Marat Gilfanov}
\affiliation{Space Research Institute, Russian Academy of Sciences, Profsoyuznaya ul. 84/32, Moscow, 117997, Russia}
\affiliation{Max-Planck-Institut f\"{u}r Astrophysik, Karl-Schwarzschild-Str. 1, D-85741 Garching, Germany}

\author[0000-0002-0129-806X]{K. R. Hinds}
\affiliation{\ljmu} 

\author[0009-0004-3067-2227]{Dan Jarvis}
\affiliation{Astrophysics Research Cluster, School of Mathematical and Physical Sciences, University of Sheffield, Sheffield S3 7RH, UK}

\author[0000-0002-5619-4938]{Mansi M. Kasliwal}
\affil{\caltech} 

\author[0000-0002-7866-4531]{Chang~Liu}
\affiliation{Department of Physics and Astronomy, Northwestern University, 2145 Sheridan Rd, Evanston, IL 60208, USA}
\affil{Center for Interdisciplinary Exploration and Research in Astrophysics (CIERA), Northwestern University, 1800 Sherman Ave, Evanston, IL 60201, USA}

\author[0000-0002-3464-0642]{Joseph D.~Lyman}
\affiliation{Department of Physics, University of Warwick, Coventry CV4 7AL, UK}

\author[0000-0002-8532-9395]{Frank J. Masci}
\affiliation{IPAC, California Institute of Technology, 1200 E. California
             Blvd, Pasadena, CA 91125, USA}

\author[0000-0002-8850-3627]{Thomas A. Prince}
\affiliation{\caltech}

\author[0000-0002-7252-5485]{Vikram Ravi}
\affiliation{\caltech} 

\author[0000-0003-0427-8387]{R. Michael Rich}
\affiliation{Department of Physics and Astronomy, UCLA, 430 Portola Plaza, Box 951547, Los Angeles, CA 90095-1547, USA}

\author{Reed Riddle}
\affiliation{\coo} 

\author[0009-0003-2780-704X]{Jason Sevilla}
\affiliation{Department of Astronomy, Cornell University, Ithaca, NY 14853, USA}

\author[0000-0001-7062-9726]{Roger Smith}
\affiliation{\coo} 

\author[0000-0003-1546-6615]{Jesper Sollerman}
\affiliation{Department of Astronomy, The Oskar Klein Center, Stockholm University, AlbaNova, 10691 Stockholm, Sweden}

\author[0000-0001-8426-5732]{Jean J. Somalwar}
\affiliation{\caltech} 

\author[0000-0002-6428-2700]{Gokul P. Srinivasaragavan}
\affiliation{Department of Astronomy, University of Maryland, College Park, MD 20742, USA}
\affiliation{Joint Space-Science Institute, University of Maryland, College Park, MD 20742, USA}
 \affiliation{Astrophysics Science Division, NASA Goddard Space Flight Center, 8800 Greenbelt Rd, Greenbelt, MD 20771, USA}

\author{Rashid Sunyaev}
\affiliation{Space Research Institute, Russian Academy of Sciences, Profsoyuznaya ul. 84/32, Moscow, 117997, Russia}
\affiliation{Max-Planck-Institut f\"{u}r Astrophysik, Karl-Schwarzschild-Str. 1, D-85741 Garching, Germany}

\author[0009-0000-4044-8799]{Jada L.~Vail}
\affiliation{Department of Astronomy, Cornell University, Ithaca, NY 14853, USA}

\author[0000-0003-0733-2916]{Jacob L.~Wise}
\affiliation{\ljmu} 

\author[0009-0000-4440-155X]{Sol Bin Yun}
\affiliation{\caltech} 

\begin{abstract}

We present the optical discovery and multiwavelength follow-up observations of AT\,2024kmq, a likely tidal disruption event (TDE) associated with a supermassive ($M_{\rm BH}\sim 10^{8}\,M_\odot$) black hole in a massive galaxy at $z=0.192$. 
The optical light curve of AT\,2024kmq exhibits two distinct peaks: an early fast (timescale 1\,d) and luminous ($M\approx-20\,$mag) red peak, then a slower (timescale 1\,month) blue peak with a higher optical luminosity ($M\approx-22\,$mag) and featureless optical spectra. The second component is similar to the spectroscopic class of ``featureless TDEs'' in the literature, and during this second component we detect highly variable, luminous ($L_X\approx 10^{44}\,$erg\,s$^{-1}$), and hard ($f_\nu \propto \nu^{-1.5}$) X-ray emission. 
Luminous ($10^{29}\,$erg\,s$^{-1}$\,Hz$^{-1}$ at 10\,GHz) but unchanging radio emission likely arises from an underlying active galactic nucleus. The luminosity, timescale, and color of the early red optical peak can be explained by synchrotron emission, or alternatively by thermal emission from material at a large radius ($R\approx\mathrm{few}\times10^{15}\,$cm). Possible physical origins for this early red component include an off-axis relativistic jet, and shocks from self-intersecting debris leading to the formation of the accretion disk. Late-time radio observations will help distinguish between the two possibilities.

\end{abstract}
\keywords{
Tidal disruption (1696);
X-ray transient sources (1852); 
Supermassive black holes (1663);
Time domain astronomy (2109); 
High energy astrophysics (739); 
Accretion (14)
}

\vspace{1em}

\section{Introduction}

The tidal disruption and accretion of a star by a black hole can produce a luminous electromagnetic transient (\citealt{Hills1975, Rees1988}; see \citealt{Komossa2015} and \citealt{Gezari2021} for recent reviews). 
Wide-field time-domain surveys such as the Zwicky Transient Facility (ZTF; \citealt{Graham2019,Bellm2019b}) in the optical band have enabled large-scale systematic studies of tidal disruption events (TDEs), with over 100 discovered to date, and the delineation of several spectroscopic classes \citep{vanVelzen2021,Gezari2021,Hammerstein2023,Yao2023}. Major open questions in the TDE field include how the accretion disk forms following the disrupted star's return to pericenter \citep{Lodato2015,Bonnerot2021_review}, and under what circumstances the accretion process leads to the launch of a relativistic jet \citep{DeColle2020}. 

Typically, the optical emission observed in TDEs is blue (thermal with $T_{\mathrm{BB}}\sim10^{4}\,$K) and evolves on a timescale of weeks to months \citep{Gezari2021}. The powering mechanism for the optical emission is uncertain: possibilities include accretion onto the black hole (e.g., \citealt{Strubbe2009}), shocks from self-intersecting debris streams (e.g., \citealt{Piran2015}), and reprocessing of soft X-rays (e.g., \citealt{Roth2016}). 
However, one TDE was recently discovered via rapidly fading, red, nonthermal optical emission (AT\,2022cmc; \citealt{Andreoni2022}). The optical emission---alongside luminous X-ray and radio emission---has been interpreted as synchrotron radiation from the afterglow of an on-axis jet \citep{Andreoni2022,Pasham2023}. 

AT\,2022cmc is one of only four on-axis jetted TDEs discovered to date (see \citealt{DeColle2020} for a review); the other three \citep{Bloom2011,Burrows2011,Zauderer2011,Levan2011,Cenko2012,Brown2015} were identified as high-energy transients by the Burst Alert Telescope \citep{Barthelmy2005} on the {\it Neil Gehrels Swift Observatory} (\emph{Swift}; \citealt{Gehrels2004}). Two of the on-axis jetted TDEs, including AT\,2022cmc, had optical properties closely resembling the rare class of ``featureless TDEs,'' which are characterized by high optical luminosities ($-23\lesssim M_{\rm peak}\lesssim-21\,$mag) and featureless optical spectra \citep{Hammerstein2023}.
The observed association led \citet{Andreoni2022} to suggest a link between featureless TDEs and the presence of a relativistic jet. 

In this paper we present AT\,2024kmq, a second TDE identified on the basis of early fast red optical emission, which also had a longer-lasting thermal component characteristic of featureless TDEs. The X-ray luminosity of AT\,2024kmq is intermediate between ``ordinary'' optical TDEs and on-axis jetted TDEs, and there is no clear radio emission from the transient itself, although deep constraints are difficult owing to luminous radio emission that is likely from the active galactic nucleus (AGN) of the host galaxy. We present the discovery and observations in Section~\ref{sec:obs}, basic inferences and comparisons to known TDEs in Section~\ref{sec:inferences}, modeling of the first optical peak in Section~\ref{sec:1st_peak}, and summarize in Section~\ref{sec:summary}. This paper focuses on data obtained of this transient prior to October 2024, when observations became limited by a Sun constraint. A subsequent paper (Yao et al. 2025) will focus on our late-time observations. 


We adopt a standard $\Lambda$CDM cosmology with matter density $\Omega_{\rm M} = 0.31$ 
and the Hubble constant $H_0=67.7\,{\rm km\,s^{-1}\,Mpc^{-1}}$ \citep{Planck2020}. Unless otherwise noted, uncertainties are reported using 68\% confidence intervals and upper limits are reported at 3$\sigma$.

\section{Observations and Analysis} \label{sec:obs}

\subsection{Transient Identification and TDE Classification}
\label{sec:discovery}

AT\,2024kmq was detected as ZTF24aapvieu on\footnote{All times given in UTC.} 2024 June 1 05:21:18 at $g = 19.93\pm0.08$ mag\footnote{All magnitudes given in AB.} at $\alpha = 12^{\mathrm{h}}02^{\mathrm{m}}37\fs26$, $\delta = +35^{\circ}23^{\prime}35\farcs 5$ (J2000) as part of the Zwicky Transient Facility \citep{Graham2019,Bellm2019b} high-cadence partnership survey \citep{Bellm2019-sched} with the 
48-inch Samuel Oschin Schmidt telescope at Palomar Observatory (P48). The ZTF observing system is described in \citet{Dekany2020} and the data processing pipeline is described in \citet{Masci2019}. The identification of AT\,2024kmq made use of machine learning-based real-bogus classifiers \citep{Mahabal2019,Duev2019} and a star-galaxy separator \citep{Tachibana2018}. AT\,2024kmq was also detected by the Gravitational-wave Optical Transient Observer (GOTO; \citealt{Dyer2024}) one day prior to the first ZTF detection, at 2024 May 30 21:16:09. 

\begin{figure*}[htbp!]
\centering
    \includegraphics[width=1.0\textwidth]{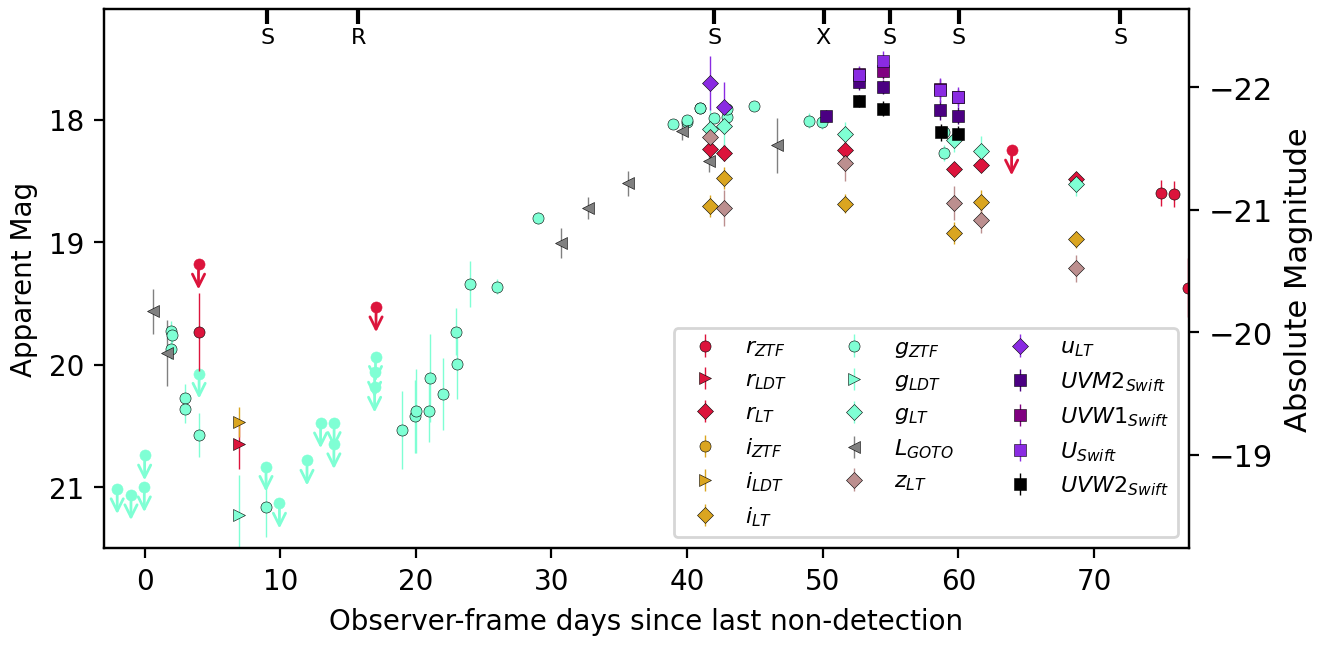}
    \caption{Select optical and UV photometry of AT\,2024kmq (corrected for Milky Way extinction) with epochs of key follow-up observations marked: S for spectroscopy, R for the first radio observation, and X for the first X-ray observation. The absolute magnitude has a cosmological correction of $2.5\log_{10}(1+z)$ applied. Upper limits (5-$\sigma$) are marked with downward-facing arrows.} 
    \label{fig:opt_lc}
\end{figure*}

Upon the first ZTF detection on 2024 June 1, AT\,2024kmq passed a filter designed to identify rapidly brightening luminous transients (e.g., \citealt{Ho2020}). The rise rate was 0.5\,mag\,d$^{-1}$ in $g$ band from a non-detection two nights prior. The ZTF transient position was 0.5$^{\prime\prime}$ offset\footnote{Subsequent deeper optical imaging suggests that the transient is in fact nuclear; see Section~\ref{sec:opt-phot}.} from the nucleus of the catalogued galaxy Sloan Digital Sky Survey (SDSS) J120237.22+352335.3. From the SDSS photometric redshift of $z_\mathrm{ph}=0.231 \pm 0.014$ \citep{Beck2016}, the implied peak luminosity of the transient was $\approx-20.4\,$mag. Over the next two nights, daily ZTF data showed fading of almost 1\,mag in $g$ band, implying that this was a short-duration transient. 
The optical light curve is shown in Figure~\ref{fig:opt_lc}.

To confirm the rapid decline, we triggered imaging observations (described in more detail in Section~\ref{sec:opt-phot}) using
the Spectral Energy Distribution Machine (SEDM; \citealt{Blagorodnova2018,Rigault2019,Kim2022}) on the automated 60-inch telescope at Palomar Observatory (P60; \citealt{Cenko2006}), and the Large Monolithic Imager (LMI) on the Lowell Discovery Telescope (LDT). Imaging obtained on 2024 June 6 showed fast fading from the ZTF detections. In addition, the LDT imaging revealed red colors ($g-r=0.6\,$mag, corrected for $A_V=0.054$ of Milky Way extinction; \citealt{Schlafly2011}), distinguishing AT\,2024kmq from the typically blue colors of extragalactic day-timescale transients \citep{Drout2014,Pursiainen2018,Ho2023}.

To measure the transient redshift, on 2024 June 7 we triggered a Target-of-Opportunity (ToO) program\footnote{Program ID GN-2024A-Q-127, PI Ho} using the Gemini Multi-Object Spectrograph (GMOS; \citealt{Hook2004}) on Gemini North (more details in Section~\ref{subsec:opt-spec}). We also reported the event to the Transient Name Server \citep{Vail2024}. 
As discussed in \cite{Sevilla2024}, the GMOS spectrum was dominated by the host galaxy. Detections of \ion{Na}{I}, \ion{Mg}{I}, H$\alpha$, and H$\beta$ absorption lines yielded a spectroscopic redshift of $z=0.192$ and an implied absolute peak magnitude (with a cosmological correction; $M=m_\mathrm{obs}-5\log_{10}(D_L/10\,\mathrm{pc}) + 2.5\log_{10}(1+z)$) of $M_g = -19.97\pm0.08$\,mag. The lack of any emission lines in the host galaxy was unusual for a fast extragalactic transient \citep{Wiseman2020} but not unprecedented \citep{Nicholl2023}. Due to the confirmed high optical luminosity, we triggered a radio ToO program using the Very Large Array; radio observations are described in more detail in Section~\ref{sec:radio}. 

Starting on 2024 June 18, the ZTF $g$-band light curve showed significant rebrightening over a timescale of weeks, motivating us to pursue a multiwavelength observing campaign. 
We classify AT\,2024kmq as a TDE based on the persistent blue color and hot photospheric temperature of the second (primary) peak; the featureless optical spectra; and the mass and color of its host galaxy (\S\ref{subsubsec:uvot}), which are reminiscent of the TDE-featureless spectral subtype \citep{Hammerstein2023}. 

Throughout the paper we use the time of the last ZTF non-detection ($t_0=60460.24495$ MJD) as our reference epoch, which is 0.64\,d prior to the first GOTO detection. Follow-up observations were coordinated using the SkyPortal \citep{vanderWalt2019,Coughlin2023} platform.


\subsection{Ground-based Optical Photometry}
\label{sec:opt-phot}

Our optical photometry of AT\,2024kmq is summarized in Table~\ref{tab:optical-photometry} in Appendix~\ref{sec:appendix}.
We obtained six epochs of multi-band Liverpool Telescope (LT; \citealt{Steele2004}) IO:O photometry of AT\,2024kmq. We performed astrometric alignment on images that had been reduced using the standard LT pipeline. Image subtraction was conducted using the Pan-STARRS1 Surveys (PS1; \citealt{Chambers2016}) imaging catalog as a reference and a custom IDL routine (the PS1 image was convolved to match the PSF of the LT image, then subtracted). Transient photometry was performed using seeing-matched aperture photometry fixed at the transient location and calibrated relative to a set of SDSS secondary standard stars in the field (as measured from the unsubtracted images). We performed photometry in the same way for the one epoch of LDT/LMI imaging and the four epochs of P60/SEDM imaging. We added (0.3\,mag, 0.2\,mag, 0.1\,mag) of systematic uncertainty in quadrature to the ($g$, $r$, $i$)-band LDT/LMI data. Point-spread-function (PSF)-fit forced photometry was performed on archived difference images from the ZTF survey using the ZTF forced-photometry service \citep{Masci2023}. 

The Gravitational-wave Optical Transient Observer (GOTO; \citealt{Dyer2024}) data were observed as part of routine sky-survey operations in the GOTO-L band ($\sim400$--$700$\,nm) and reduced in real-time via an internal pipeline (Lyman et al. in prep). Forced photometry was performed in small apertures on difference images at the location of AT\,2024kmq via the GOTO light curve service (Jarvis et al. in prep). These small-aperture measurements were corrected for missing flux using a local measurement of the aperture correction from nearby bright stars. Image-level zeropoints were calculated using ATLAS-REFCAT2 \citep{Tonry2018cat}, with a position-dependent correction to the image-level value applied, again using bright stars in the field. 

The position of the optical transient as measured by LDT is consistent with the host-galaxy nucleus. More specifically, in the $r$ band, which has the cleanest subtraction, the offset of the transient from the host-galaxy nucleus is $[0.045, 0.137]^{\prime\prime}\pm[0.114, 0.132]^{\prime\prime}$ in the [E, N] direction. The uncertainty in the position is the root-mean-square astrometric scatter in the offsets of the other objects in the image. We note that at the redshift of AT\,2024kmq, $0.1^{\prime\prime}$ corresponds to 0.3\,kpc. 

Finally, we searched for historical flaring at the transient position. Forced photometry on ZTF images\footnote{The data is available at IPAC: \dataset[10.26131/IRSA539]{\doi{10.26131/IRSA539}}.} \citep{Masci2023} extending back to the start of the survey did not show any clear transient event. In the AllWISE catalog, the source is categorized as likely non-variable in the W1 and W2 filters \citep{Cutri2014}.

\subsection{Optical Spectroscopy}
\label{subsec:opt-spec}

\begin{figure}[htbp!]
\centering
    \includegraphics[width=\columnwidth]{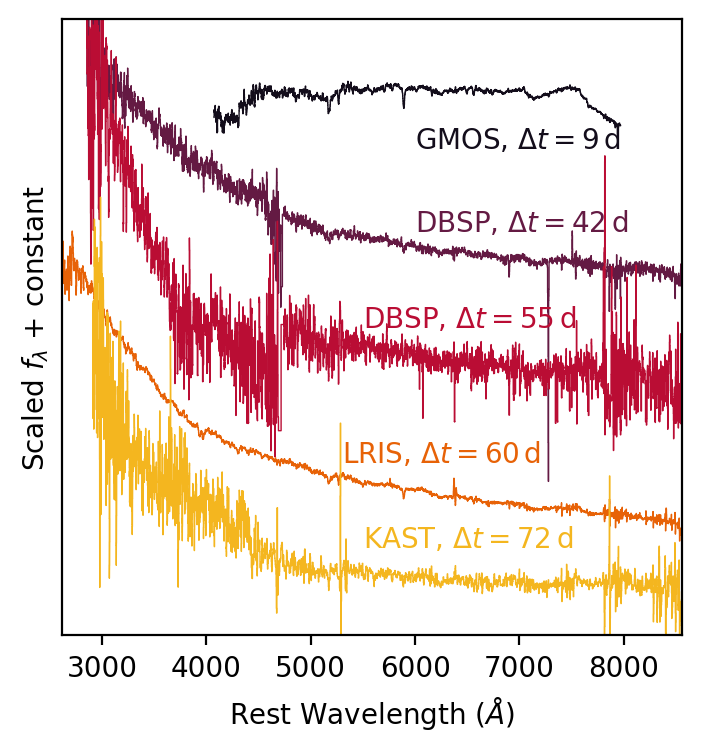}
    \caption{The optical spectra of AT\,2024kmq, binned using 3--6\,\AA\ bins. Observer-frame epoch is given with respect to $t_0$ as defined in Section~\ref{sec:discovery}. No host-galaxy subtraction has been performed, so the first spectrum is dominated by features from the host galaxy. 
    \label{fig:opt-spec}}
\end{figure}

We obtained five optical spectra of AT\,2024kmq. The epochs of spectroscopic observations are marked with
`S' in Figure~\ref{fig:opt_lc},
and observation details are provided in Table~\ref{tab:spectra}. The spectral sequence is shown in Figure~\ref{fig:opt-spec}. 

\begin{deluxetable}{lrrr}[htb!]
\tablecaption{Spectroscopic observations of AT\,2024kmq. Observer-frame epochs given since $t_0$ as defined in \S\ref{sec:discovery}.
\label{tab:spectra}} 
\tablewidth{0pt} 
\tablehead{\colhead{Start Date} & \colhead{$\Delta t$} & \colhead{Tel.+Instr.} & \colhead{Exp. Time} \\ 
\colhead{(MJD)} & \colhead{(d)} & & \colhead{(s)}} 
\startdata 
60469.3026 & 9 & Gemini+GMOS & 450  \\
60502.2383 & 42 & P200+DBSP & 900  \\
60515.1906 & 55 & P200+DBSP & 600  \\
60520.2666 & 60 & Keck1+LRIS & 605 \\
60532.1814 & 72 & Shane+KAST & 720 \\
\enddata 
\end{deluxetable}

The first spectrum was obtained using GMOS with the R400 grating and a 1\arcsec-wide slit, which gives a resolution of 959 at the blaze wavelength of 764\,nm.
We reduced the GMOS spectrum using the \texttt{PypeIt} package \citep{pypeit:joss_pub}. 
We obtained two spectra using the Double Beam Spectrograph (DBSP; \citealt{Oke1982}) on the 200-inch Hale telescope at Palomar Observatory. We used the 600/4000 grating for the blue arm and the 316/7500 grating for the red arm, covering 3200--10000 AA at an average resolution of $\sim1000$. The slit was oriented along the parallactic angle. We reduced the data using the P200/DBSP pipeline described in \citet{Roberson2022}. We obtained one spectrum using the Low Resolution Imaging Spectrometer (LRIS; \citealt{Oke1995}) on the Keck~I 10\,m telescope and a final spectrum using the Kast spectrograph on the Shane 3\,m telescope at Lick observatory. For the LRIS observation we used a 1$^{\prime\prime}$-wide slit, a 400/8500 grating (red-side CCD), and a 400/3400 grism (blue-side CCD). For the Kast observation we used the 300/7500 grating (red-side CCD, 2.55 Å/pixel dispersion) and 600/4310 grism (blue-side CCD, 1.02 Å/pixel dispersion), with a slit width of $1.5^{\prime\prime}$.
The Keck/LRIS pipeline \texttt{Lpipe} is described in \citet{Perley2019lpipe}, and the KAST spectrum was reduced using the  UCSC Spectral Pipeline \citep{Siebert2019}.

To search for broad features from the transient, we scaled and subtracted a model spectrum of the host galaxy (Section~\ref{sec:host}) from the early Gemini spectrum (corresponding to the early red component) as well as from the later LRIS spectrum (corresponding to the blue component). We did not identify any clear features from the transient itself. 

\subsection{Swift}

AT\,2024kmq was observed by the Ultra-Violet/Optical Telescope (UVOT; \citealt{Roming2005}) and the X-Ray Telescope (XRT; \citealt{Burrows2005}) on board the \emph{Swift} for five epochs (total exposure time: 7.4\,ks) from 2024 July 19 to 2024 July 29. 

\subsubsection{UVOT}\label{subsubsec:uvot}
The UVOT flux of AT\,2024kmq was measured with \texttt{uvotsource}, using a circular source region with $r_{\rm src} = 5^{\prime\prime}$. The background flux was measured using four nearby circular source-free regions with $r_{\rm bkg}=6^{\prime\prime}$. We estimated the host galaxy flux by fitting a host galaxy synthesis model on archival photometry from GALEX, SDSS, and WISE (more details in Section~\ref{sec:host}). The best-fit model gives 
observed host-galaxy magnitudes of $uvw2=25.43$\,mag, $uvm2=26.02$\,mag, $uvw1=23.75$\,mag, $U = 21.58$\,mag, $B = 19.94$\,mag and $V = 18.43$\,mag. 
Considering the relative significant host contribution in $B$ and $V$ bands and the uncertainties in the host spectral energy distribution (SED) model, we excluded these two bands from our analysis.
The host-subtracted UVOT photometry is given in Table~\ref{tab:optical-photometry} in Appendix~\ref{sec:appendix}.


\subsubsection{XRT}
\label{subsubsec:xrt}

We processed the XRT data using \texttt{HEASoft} version 6-33.2. The source was detected 
in all five obsIDs. We identified no significant hardness evolution across the obsIDs. We created an X-ray light curve per good time interval (GTI), then grouped adjacent GTIs within a single obsID together if the change in count rate was less than $1\sigma$, resulting in six time bins. 

To generate the time-averaged source and background XRT spectra, we stacked the cleaned event files with \texttt{xselect}, and filtered using a source region with $r_{\rm src}=40^{\prime\prime}$, and eight background regions with $r_{\rm bkg}=45^{\prime\prime}$ evenly spaced at $100^{\prime\prime}$ from AT\,2024kmq. The source spectrum was grouped with \texttt{ftgrouppha} using the optimal binning scheme developed by \citet{Kaastra2016}, requiring at least one count per bin. 

\begin{figure}[htbp]
    \includegraphics[width=\columnwidth]{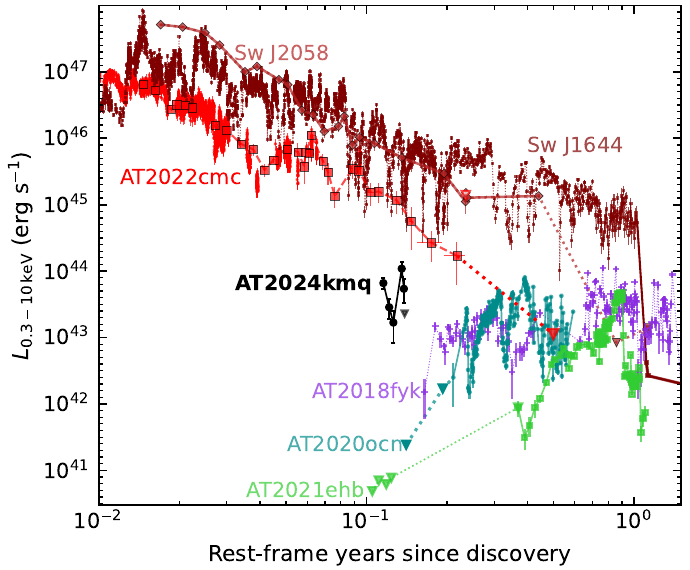}
    \caption{The X-ray light curve of AT\,2024kmq compared with X-ray light curves of TDEs that exhibit significant hard X-ray emission \citep{Pasham2015, Mangano2016, Eftekhari2018, Wevers2021, Yao2022, Yao2024, Eftekhari2024, Pasham2024_20ocn}. \label{fig:xray_lc}}
\end{figure}

Using $C$-statistics \citep{Cash1979}, we modeled the spectrum with an absorbed power law (\texttt{tbabs*zashift*powerlaw} in \texttt{xspec}). The redshift was fixed at $z=0.192$ and the hydrogen-equivalent column density $N_{\rm H}$ was fixed at the Galactic value of $1.77\times 10^{20}\,{\rm cm^{-2}}$ \citep{Willingale2013}. The best-fit model yielded a power-law photon index of $\Gamma=2.45\pm0.20$, ${\rm cstat/dof}=11.51/21$, and an observed 0.3--10\,keV flux of $4.37_{-0.51}^{+0.58}\times 10^{-13}\,{\rm erg\,s^{-1}\,cm^{-2}}$. The 0.3--10\,keV net count rate to flux conversion factor was found to be $3.86\times 10^{-11}\,{\rm erg\,count^{-1}\,cm^{-2}}$. 
A table of XRT observations is given in Table~\ref{tab:xrt} and the X-ray light curve is shown in Figure~\ref{fig:xray_lc}.

\begin{deluxetable}{crcc}[htbp!]
\tablecaption{\swift/XRT observations of AT\,2024kmq.\label{tab:xrt}}
\tablehead{
    \colhead{MJD} 
    & \colhead{Exp.}
    & \colhead{$f_{\rm X}$}
    & \colhead{$L_{\rm X}$}
    \\
    \colhead{}
    & \colhead{(s)} 
    & \colhead{($10^{-13}\,{\rm erg\,s^{-1}\,cm^{-2}}$)}
    & \colhead{($10^{43}\,{\rm erg\,s^{-1}}$)}
    }
\startdata
60510.3208 & 1988 & $6.28 \pm 1.23$ & $6.59 \pm 1.29$ \\
60512.9092 & 1903 & $2.72 \pm 0.89$ & $2.86 \pm 0.94$ \\
60514.8491 & 1608 & $1.60 \pm 0.81$ & $1.68 \pm 0.85$ \\
60518.9262 & 759 & $10.36 \pm 2.60$ & $10.87 \pm 2.72$ \\
60520.0460 & 705 & $5.16 \pm 2.02$ & $5.41 \pm 2.12$ \\
60520.4383 & 463 & $< 2.18$ & $< 2.29$ \\
 \enddata
\tablecomments{$f_{\rm X}$ and $L_{\rm X}$ are given in the 0.3--10\,keV energy range. }
\end{deluxetable}

\subsection{eROSITA}
\label{sec:erosita}

The position of AT\,2024kmq was observed by the eROSITA telescope \citep{Predehl2021} on board the \textit{Spektrum-Roentgen-Gamma} (\srg) satellite \citep{Sunyaev2021}. The position was observed on four epochs from 2020--2022, each separated by 6\,months, with the first observation held on 2020 May 27 (MJD 58996.43). The 0.2--2.3\,keV upper limit is $\sim 4\times 10^{-14}\,{\rm erg\,s^{-1}\,cm^{-2}}$ in individual scans, and $\sim 1.5\times 10^{-14}\,{\rm erg\,s^{-1}\,cm^{-2}}$ in the combined data of all four observations. 
The latter upper limit corresponds to $L_X<1.6\times 10^{42}\,$erg\,s$^{-1}$. Therefore, the X-ray emission observed by XRT is clearly dominated by the transient.


\subsection{Radio}
\label{sec:radio}

\subsubsection{VLA}

We obtained four epochs of observations of AT\,2024kmq\footnote{Programs 23B-138 (PI A.~Ho) \& 24A-494 (PI Y.~Yao).} from MJD 60475.98 to 60565.03 ($15.09$--$140.14~$days post discovery) with the Karl G. Jansky Very Large Array (VLA; \citealt{Perley2011}), at mean frequencies of 6, 10, and 15~GHz.
3C286 was used as the flux density and bandpass calibrator and J1215+3448 as the complex gain calibrator and all observations were taken in B configuration.
The data were calibrated and imaged under the Science Ready Data Products (SRDP) initiative\footnote{https://science.nrao.edu/srdp}, which uses the Common Astronomy Software Applications (CASA; \citealt{McMullin2007}) VLA Calibration Pipeline\footnote{https://science.nrao.edu/facilities/vla/data-processing/pipeline} and VLA Imaging Pipeline\footnote{https://science.nrao.edu/facilities/vla/data-processing/pipeline/VIPL}. We downloaded the regularly calibrated continuum images for each epoch and mean frequency.

There is a clear radio source detected at the position of the host-galaxy nucleus. To determine the flux density and shape of the source, 
we used the \texttt{pwkit/imtool} program \citep{2017ascl.soft04001W}. 
We did not find a significant difference in the inferred flux density when modeling the source as point-like or extended, so for consistency we measured the flux density for all epochs assuming a point-like source.
The results of this analysis are presented in Table~\ref{tab:radio} in Appendix~\ref{sec:appendix} and in Figure~\ref{fig:radio-sed}. 
Due to the limited variability and the archival detection of extended radio emission (Section~\ref{sec:radio_archive}), we conclude that the radio emission likely arises from an AGN in the host galaxy, and is unassociated with the transient (more details in Section~\ref{sec:host}).

We note that a radio source at the position of AT\,2024kmq was detected by the Arcminute Microkelvin Imager - Large Array (AMI-LA) on 2024 July 25 ($55~$days post detection) at a mean frequency of 15.5~GHz and a flux density of $f_\nu = 0.171 \pm 0.020~$mJy \citep{2024TNSAN.205....1L}. This is $\sim 2$ times brighter than the VLA observations encompassing the AMI-LA observations at $45$ and $75$ days. The discrepancy could potentially be due to the source being extended (Section~\ref{sec:radio-properties}).

For the remainder of the paper we adopt the measured point-source brightness of the AGN ($\approx0.1$\,mJy at 10\,GHz) as the upper limit on the brightness of the AT\,2024kmq transient emission.

\subsubsection{Archival Radio Imaging}
\label{sec:radio_archive}

We utilized the LOw-Frequency ARray (LOFAR) Two-metre Sky Survey (LoTSS) Data Release 2 cutout API to generate a cutout\footnote{The cutout header does not have an observation date, but all of the observations for LoTSS DR2 were taken before the first optical detection of AT\,2024kmq.} at the position of AT\,2024kmq \citep{Shimwell2022}. There is a clearly detected and extended source at the position of AT\,2024kmq, coincident with the VLA radio source. \texttt{imtool} prefers an extended profile (approximate size of $14.1\arcsec \times 8.6\arcsec$, compared to a beam size of $6\arcsec \times 6\arcsec$) for this source, and we measure a flux density of $f_\nu = 6.74\pm 0.22~$mJy (compared to only $f_\nu = 2.34\pm 0.62~$mJy when the shape is forced to be point-like) at a mean frequency of $150~$MHz. Similarly, we utilized the CSIRO Australian SKA Pathfinder (ASKAP) Science Data Archive (CASDA) Cutout Service to generate a Rapid ASKAP Continuum Survey-Mid (RACS-Mid) cutout from an observation taken on 2020 December 24 \citep{2023PASA...40...34D}. At the position of AT\,2024kmq we measured a flux density of $f_\nu = 1.09\pm 0.20~$mJy at a mean frequency of $1.4~$GHz. The extended size inferred from the LoTSS imaging corresponds to $\approx50\,$kpc at the distance of AT\,2024kmq, which is typical for radio-loud AGN \citep{Hardcastle2019}.

\begin{figure}[htbp]
    \centering \includegraphics[width=\columnwidth]
    {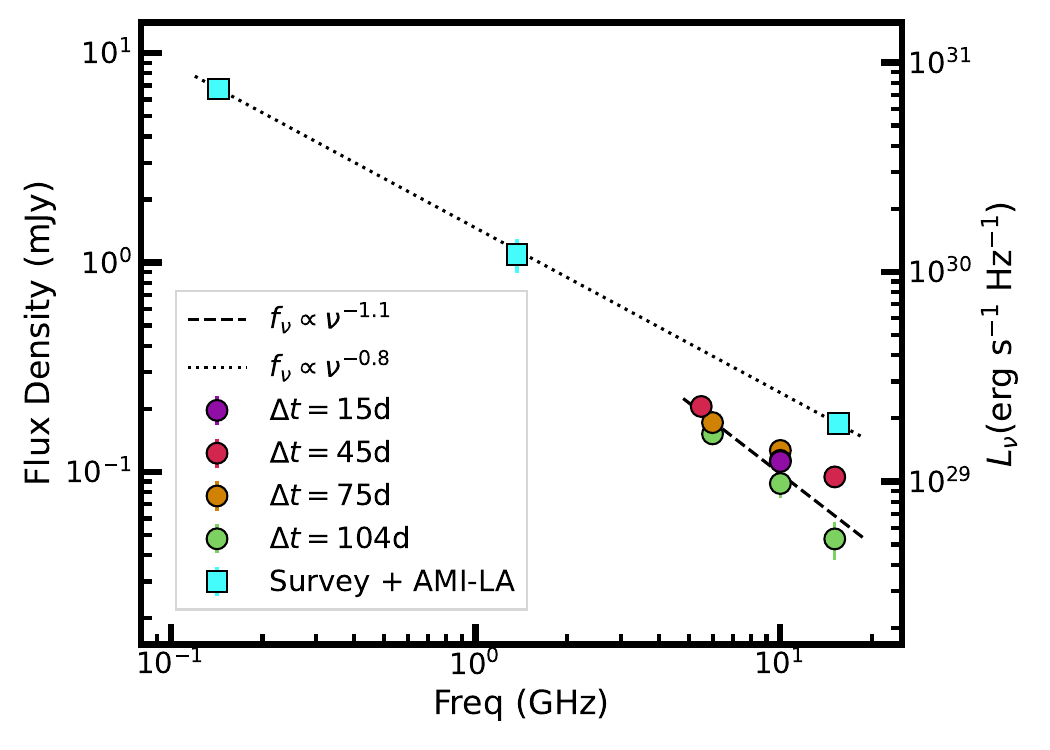}
    \caption{The SED of a radio source consistent with the nucleus of the AT\,2024kmq host galaxy. Circles mark VLA data presented in this paper. Squares mark survey data (LOFAR and ASKAP) from prior to the optical transient, as well as AMI-LA data from $\Delta t=55\,$d.}
    \label{fig:radio-sed}
\end{figure}

\section{Basic Properties and Inferences}
\label{sec:inferences}

\subsection{Host Galaxy}
\label{sec:host}


As mentioned in Section~\ref{subsubsec:uvot}, we fit a host-galaxy model to the GALEX, SDSS, and WISE photometry following the same procedures adopted by \citet{vanVelzen2021} and \citet{Hammerstein2023}. Specifically, we use Prospector \citep{Johnson2021}, the Markov Chain Monte Carlo sampler \texttt{emcee} \citep{Foreman-Mackey2013}, and the Flexible Stellar Population Synthesis (FSPS) models \citep{Conroy2009}. The best-fit host-galaxy properties are given in Table~\ref{tab:host}. 

\begin{figure}[htbp]
    \centering
    \includegraphics[width=\columnwidth]{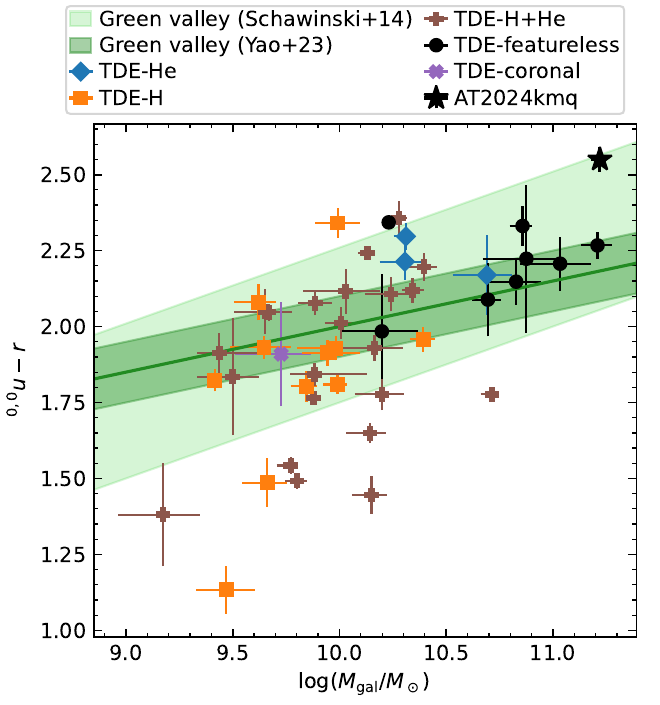}
    \caption{The host galaxy of AT\,2024kmq on the $^{0,0}u-r$ vs. $M_{\rm gal}$ diagram, compared with published ZTF TDE samples color-coded by the optical spectral subtype (\citealt{vanVelzen2021, Hammerstein2023, Yao2023}). Green bands mark the green valley defined by \citet{Schawinski2014} and \citet{Yao2023}.
    \label{fig:Mgal_umr}}
\end{figure}

In Figure~\ref{fig:Mgal_umr}, we compare the stellar mass and the Galactic extinction-corrected, synthetic
rest-frame $u-r$ color of the host galaxy of AT\,2024kmq to those of known ZTF TDEs. 
The comparison sample of 45 TDEs is constructed by combining all events presented by \citet{vanVelzen2021}, \citet{Hammerstein2023}, and \citet{Yao2023} that have a known optical spectral subtype. The host-galaxy properties were measured in the same fashion (Prospector and FSPS).  
As can be seen, the host galaxy of AT\,2024kmq is most similar to previously known featureless TDEs, which are found to be preferentially hosted by high-mass galaxies. We note that the 95\% confidence upper limit on the host-galaxy mass of AT2022cmc is $\log(M/M_\odot)<11.2$ \citep{Andreoni2022}.


\begin{deluxetable}{ll}
\tablecolumns{2}
\tablecaption{Best-fit host galaxy properties of AT\,2024kmq. \label{tab:host}}
\tablehead{\colhead{Parameter} & \colhead{Value}}
\startdata
$\log (M_{\rm gal}/M_\odot)$ & $11.22_{-0.04}^{+0.03}$\\
$^{0,0}u-r$ (mag) & $2.55\pm0.04$ \\
$E_{B-V, {\rm host}}$ (mag) & $0.06_{-0.04}^{+0.08}$ \\
$t_{\rm age}$ (Gyr) & $11.21_{-1.52}^{+0.97}$\\
$\tau_{\mathrm{sfh}}$ (Gyr) & $0.30_{-0.16}^{+0.37}$ \\
$\log (Z/Z_\odot)$ & $-0.10_{-0.10}^{+0.08}$ \\
\enddata
\end{deluxetable}

The $M_{\rm gal}$--$M_{\rm BH}$ scaling relation presented in \citet{Greene2020}, which was derived using all galaxies with dynamical measurements and upper limits on $M_{\rm BH}$, implies a black hole mass for AT\,2024kmq of ${\rm log}(M_{\rm BH}/M_\odot)=8.63\pm0.83$. The $M_{\rm gal}$--$M_{\rm BH}$ scaling relation derived in \citet{Yao2023} using 19 TDE hosts with velocity dispersion measurements implies a black hole mass of ${\rm log}(M_{\rm BH}/M_\odot)=8.54\pm0.37$. 

Using the GMOS spectrum (Section~\ref{subsec:opt-spec}), we measured the stellar velocity dispersion ($\sigma_\ast$) of the host galaxy using \texttt{pPXF} \citep{Cappellari2004, Cappellari2017}. Following the procedures adopted by \citet{Yao2023}, we obtained $\sigma_{\ast} = 175.4^{+27.0}_{-26.9}\,{\rm km\,s^{-1}}$. Using the \citet{Kormendy2013} $M_{\rm BH}$--$\sigma_\ast$ scaling relation, this implies a black hole mass of ${\rm log}(M_{\rm BH}/M_\odot)=8.24\pm0.30({\rm stat})\pm0.29({\rm sys})$. 

If the disrupted star is a Sun-like star, the Hills mass\footnote{The Hills mass is defined as the maximum black hole mass for an observable TDE where the tidal radius is outside of the event horizon radius \citep{Hills1975}.} is only above $4\times 10^8\,M_\odot$ for a highly spinning black hole with spin parameter $a_{\rm BH}\gtrsim 0.95$ (see, e.g., Fig.~4 of \citealt{leloudas2016}). If the disrupted star is a higher-mass main-sequence or evolved star, smaller values of $a_{\rm BH}$ are allowed. We note that $M_{\rm gal}$--$M_{\rm BH}$ relations do have significant scatter: early-type galaxies in \citet{Greene2020} with $M_\mathrm{gal}\sim10^{11}\,M_\odot$ have black-hole masses down to $10^{7}\,M_\odot$. 


The luminous and relatively unvarying radio emission (Section~\ref{sec:radio}) suggests that the host galaxy of AT\,2024kmq harbors an AGN, which we discuss in further detail in Section~\ref{sec:radio-properties}. Our estimate of the host-galaxy mass is not affected by excluding the WISE photometry from the SED fit. 


\subsection{Origin of Early-time Hard X-ray Emission}

In general, although a significant fraction (at least 40\%) of optically selected TDEs eventually exhibit $>10^{42}\,$erg\,s$^{-1}$ X-ray emission, luminous ($10^{44}\,$erg\,s$^{-1}$) X-ray emission at the time of optical peak is rare (e.g., \citealt{Guolo2024}). While most TDEs exhibit thermal soft X-ray emission originating from the inner accretion disk, significant non-thermal hard X-rays (with $\Gamma<3$) have so far only been observed in two small groups of TDEs (see Figure~\ref{fig:xray_lc}): those with on-axis jets and those with a late-time soft-to-hard spectral transition at ${\rm few}\times 10^2$\,days after the optical peak. 
The former group features prompt, very luminous ($\sim\!10^{47}\,{\rm erg\,s^{-1}}$) hard X-ray emission powered by internal energy dissipation within an on-axis jet \citep{Burrows2011, Yao2024}. 
The latter group includes AT\,2018fyk \citep{Wevers2021}, AT\,2021ehb \citep{Yao2022}, and AT\,2020ocn \citep{Guolo2024}, and their spectral hardening has been explained by the gradual formation of a magnetically dominated corona above the newly formed accretion disk, where inverse Compton scattering of thermal seed photons in the corona produce the power-law X-ray spectra. 

The X-ray properties of AT\,2024kmq are different from the above two groups. It developed hard X-rays at an early phase, no later than the primary optical peak, but at a luminosity that is much fainter than on-axis jetted TDEs. We consider two possibilities for the origin of its hard X-rays: a jet viewed off-axis (Section~\ref{sec:model-offaxis}), and the inner accretion disk (Section~\ref{sec:model-streams}) provided that Comptonization is efficient\footnote{In some other TDEs, power-law X-ray emission only appears at very late times when the system becomes sub-Eddington, but AT\,2024kmq might be different. As discussed in Section~\ref{sec:host}, its BH mass is likely large ($M_{\rm BH}\sim 10^8\,M_\odot$), so $L_{\rm X}\sim10^{44}\,{\rm erg\,s^{-1}}$ is sub-Eddington by $\sim 2$ orders of magnitude.}.

\subsection{Prevalence of Early Luminous Red Components in TDE Light Curves}
\label{sec:rates}

A handful of TDEs in the literature have exhibited double peaked optical light curves. For example, AT\,2023lli had a month-long ``bump'' two months prior to the peak of the optical light curve \citep{Huang2024_AT2023lli}. ASASSN-19bt exhibited a short (days- to weeks-long) luminosity spike, identified using UV+optical SED fitting \citep{Holoien2019}. However, to our knowledge, with the exception of AT\,2022cmc \citep{Andreoni2022} no other TDE has been reported to have an early component with distinct red colors like that seen in AT\,2024kmq. 

We examined the light curves of the 44 TDEs detected by ZTF at $z<0.1$ as part of the Bright Transient Survey\footnote{We required the peak brightness to be $m<19\,$mag, and imposed a ``quality cut'' in the BTS Sample Explorer \citep{Perley2020}. The quality cut is based on on the P48 coverage close to the peak of the light curve, and requires at least a $\sim1$ week cadence.} \citep{Fremling2020,Perley2020}. We selected this relatively small volume because at such nearby distances, the early component in AT\,2024kmq---which was $M<-19\,$mag for $\approx1$ week---would be a full magnitude brighter than ZTF's typical limiting magnitude of $m=20.5\,$mag. In addition, within this volume the ZTF BTS would be reasonably complete for TDEs with main peaks brighter than $M=-20\,$mag. In AT\,2024kmq, the fast red peak occurred $\approx40$ days prior to the main peak. 

We examined the 44 light curves starting from 40 days before peak. In some cases, the rise of the TDE light curve was not well sampled, or there was a long gap between the last non-detection and the first detection. However, in 21 cases an early peak of the same brightness and timescale can be ruled out. Even with $\gtrsim$week-long observing gaps, the fact that none of the remaining 23 objects have such a component detected suggests that an early peak with these characteristics (luminosity, color, timescale) is a rare phenomenon. 

\subsection{Optical Light Curve: Basic Properties}

\begin{figure*}[!ht]
    \centering
    \includegraphics[width=0.9\textwidth]{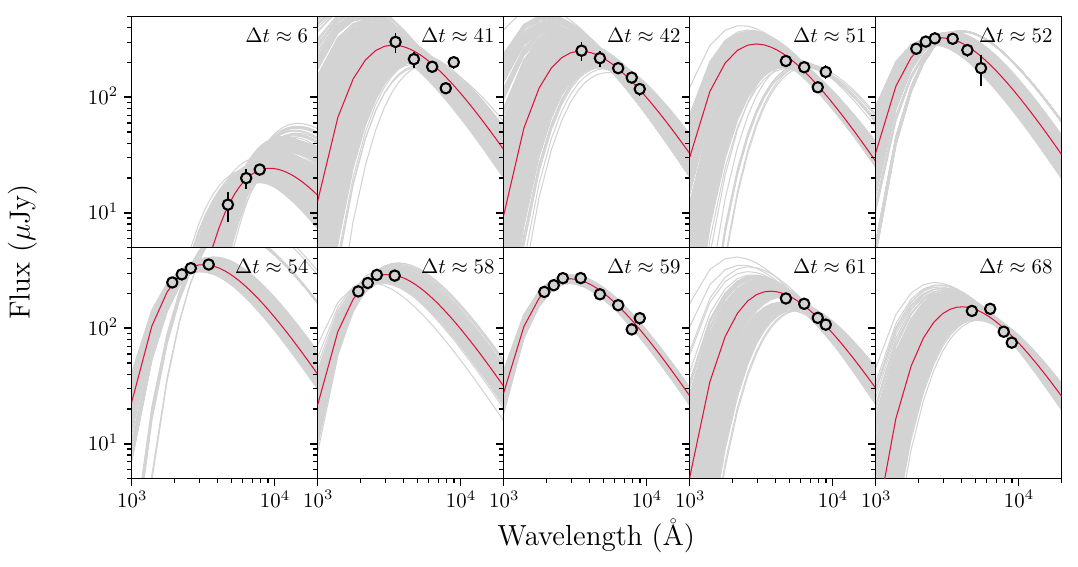}
    \caption{Blackbody fits to the LDT, LT, and/or Swift UVOT SEDs of AT\,2024kmq. Epochs are given in days from the last optical non-detection. The grey lines show each of 600 fits, and the red line shows the median value. LT observations at $\Delta t=41\,$d and $\Delta t=51\,$d were hampered by bad seeing. 
    }
    \label{fig:bbfits}
\end{figure*}

Figure~\ref{fig:opt_lc} shows two distinct components in the optical light curve: an early fast and red component, and a slower blue component. We first consider the early red component. We fit a blackbody to the LDT $gri$ imaging at $\Delta t\approx1\,$week (Figure~\ref{fig:bbfits}) and find $L_\mathrm{BB}=1.5^{+0.2}_{-0.2}\times10^{43}\,$erg\,s$^{-1}$, $T_\mathrm{BB}=5.5^{+0.9}_{-1.2}\times10^{3}\,$K, and $R_\mathrm{BB}=4.7^{+4.1}_{-1.0}\times10^{15}\,$cm. We caution that photometry in this region is challenging due to the brightness of the host galaxy---to account for this we add systematic uncertainties to the LDT measurements as described in Section~\ref{sec:opt-phot}---but the overall finding of red colors is robust. To reach this size within a week, a source starting from a much smaller size would have to expand at $v\gtrsim0.2c$. 






Another possible emission mechanism for the first peak is synchrotron radiation. For synchrotron emission we can estimate the equipartition energy $U_\mathrm{eq}$ and magnetic field strength $B_\mathrm{eq}$. The latter is, from \citet{Moffet1975},

\begin{equation}
B_\mathrm{eq} = \left( \frac{8\pi A g(\beta) L}{V} \right)^{2/7}\, ,
\end{equation}

\noindent where $A=1.586\times10^{12}$ in cgs units, $L$ is the luminosity, $V$ is the volume of the synchrotron-emitting electrons, and $g(\beta)$ is a function of the spectral index $\beta$ (defined as $f_\nu\propto \nu^{\beta}$) and frequency range ($\nu_1$ to $\nu_2$) for the power law:

\begin{equation}
    g(\beta) = \frac{2\beta+2}{2\beta+1} 
    \left[ \frac{\nu_2^{\beta+1/2}-\nu_1^{\beta+1/2}}{\nu_2^{\beta+1}-\nu_1^{\beta+1}} \right].
\end{equation}

Using the peak optical brightness, adopting a spectral index from the $gr$ observations of $f_\nu \propto \nu^{-2}$, and assuming the power law extends from $10^{13}$\,Hz to $10^{15}$\,Hz, we find $B_\mathrm{eq}\approx10\,\mathrm{G}$, which is relatively insensitive to our choices of $\nu_1$, $\nu_2$, and $\alpha$.

Next, we estimate the equipartition energy,

\begin{equation}
    U_\mathrm{eq} = 2 \frac{VB^2}{8\pi}.
\end{equation}

\noindent and find $U_\mathrm{eq}\approx 6\times10^{47}\,$erg. 

Finally, we consider the slower blue component, which is similar to that of optically discovered featureless TDEs. We construct an SED at each epoch of multi-band LT and/or Swift UVOT observations. To each SED epoch, we fit a blackbody using a Monte Carlo simulation with 600 trials (Figure~\ref{fig:bbfits}), i.e., we 
resample the observed fluxes assuming a Gaussian distribution with a standard deviation equal to the measurement uncertainty. The resulting best-fit blackbody parameters are provided in Table~\ref{tab:bbfits}. The blackbody temperatures of a $\mathrm{few}\times10^{4}\,$K are similar to those of previously observed TDEs \citep{Hammerstein2023,Yao2023}.


\begin{table}[]
    \centering
    \begin{tabular}{c|c|c|c}
    \hline\hline
       $\Delta t$  & $L$ & $T$ & $R$ \\
        (d) & ($10^{44}\,$erg\,s$^{-1}$) & ($10^4$\,K) & ($10^{15}\,$cm) \\
    \hline
    6 & $0.15^{+0.02}_{-0.02}$ & $0.55^{+0.09}_{-0.12}$ & $4.7^{+4.1}_{-1.0}$ \\
    $^*$41  & $4.6^{+3.9}_{-1.5}$ & $1.5^{+0.6}_{-0.3}$  & $3.5^{+0.9}_{-1.0}$ \\
    42 & $4.1^{+5.5}_{-1.2}$ & $1.5^{+0.8}_{-0.2}$  & $3.5^{+0.7}_{-1.2}$ \\
    $^*$51 & $5.2^{+0.9}_{-2.2}$ & $1.7^{+0.1}_{-0.5}$  & $2.9^{+1.1}_{-0.1}$ \\
    $^\dag$52 & $5.9^{+0.7}_{-0.4}$ & $1.7^{+0.3}_{-0.1}$  & $3.2^{+0.5}_{-0.7}$ \\
    $^\dag$54 & $6.1^{+0.2}_{-0.2}$ & $1.6^{+0.1}_{-0.1}$  & $3.6^{+0.4}_{-0.4}$ \\
    $^\dag$58 & $5.2^{+0.2}_{-0.2}$ & $1.7^{+0.1}_{-0.1}$  & $3.1^{+0.4}_{-0.4}$ \\
    $^\dag$59 & $5.0^{+0.2}_{-0.2}$ & $1.7^{+0.0}_{-0.0}$  & $2.8^{+0.1}_{-0.1}$ \\
    61 & $3.2^{+1.0}_{-1.2}$ & $1.4^{+0.2}_{-0.3}$  & $3.4^{+0.9}_{-0.6}$ \\
    68 & $2.1^{+0.6}_{-0.4}$ & $1.3^{+0.2}_{-0.1}$  & $3.3^{+0.4}_{-0.4}$ \\
    \hline\hline
    \end{tabular}
    \caption{Best-fit blackbody parameters for the optical and UV light curve of AT\,2024kmq. The first row corresponds to the early red peak and the remaining rows correspond to the slower blue component. $^\dag$Measurement includes \emph{Swift}/UVOT photometry. $^*$Observations impacted by poor seeing conditions.}
    \label{tab:bbfits}
\end{table}

\subsection{Radio Emission: Basic Properties}
\label{sec:radio-properties}

As discussed in Section~\ref{sec:radio}, we detect luminous radio emission at the position of AT\,2024kmq, but this radio source is likely dominated by an AGN. 
Radio AGN are known to occur preferentially in massive early-type galaxies (e.g., \citealt{Best2005}). 
From our LoTTS DR2 flux density measurement, we calculate a radio luminosity of $L_{150~{\rm MHz}}\approx 7 \times 10^{23}~{\rm W~Hz}^{-1}$ for the host of AT\,2024kmq. According to \citet{2019A&A...622A..17S}, for galaxies of a similar mass, approximately $\sim 2\%$ will be as luminous as the host of AT\,2024kmq at 150~MHz.
In addition, in galaxies as massive as the host of AT\,2024kmq it is common for the AGN to lack detected optical emission lines \citep{Kauffmann2008}, consistent with our GMOS spectrum of the host galaxy of AT\,2024kmq (Section~\ref{sec:discovery}). Also, radio-selected AGN tend not to be selected as optical or infrared AGN, and to have very low accretion rates (small Eddington ratios; \citealt{Hickox2009}). 

The AGN origin can also explain the discrepancy between the AMI-LA 15.5~GHz measurements and our VLA 15~GHz observations. The AMI-LA beam is much larger than the VLA beam, and therefore is sensitive to extended emission, which would result in a brighter source. Indeed, if we compare the LoTSS 150~MHz, RACS-Mid 1.4~GHz, and AMI-LA 15.5~GHz measurements, we find that they trace a continuous radio spectrum with spectral index $\beta \approx -0.8$ (using the convention $f_\nu \propto \nu^\beta$, Figure~\ref{fig:radio-sed}), similar to the typical values for AGN \citep{2002AJ....124..675C,2019A&A...622A..17S}. 

\citet{Alexander2020} define ``radio-loud'' TDEs to have $\nu L_\nu > 10^{40}$\,erg\,s$^{-1}$, while the radio source coincident with AT\,2024kmq is at $\approx 10^{39}\,$erg\,s$^{-1}$. Any cm-wave radio counterpart of AT\,2024kmq is at least two orders of magnitude less luminous than that of on-axis jetted TDEs at $\approx100\,$d. 
The observed spectral index of close to $F_\nu \propto \nu^{-1}$ (Figure~\ref{fig:radio-sed}) is similar to what was observed for the host galaxy of the TDE ASASSN-14li \citep{Alexander2016}. 


\section{Origin of the First Optical Peak} \label{sec:1st_peak}

We consider two possible origins for the early red peak in the optical light curve: an off-axis jet (Section~\ref{sec:model-offaxis}), and shocks from self-intersecting debris (Section~\ref{sec:model-streams}). 

\subsection{Off-axis Jet Model}
\label{sec:model-offaxis}


Our consideration of an off-axis jet model is motivated by the fact that the only other TDE with early fast red optical emission was AT\,2022cmc, widely modeled as an on-axis jetted TDE \citep{Andreoni2022,Pasham2023,Matsumoto2023,Yao2024}. The early fast red emission in AT\,2024kmq is two orders of magnitude less luminous, so we consider the possibility that the jet was viewed off-axis. The effect of viewing the jet off-axis can be seen from the prescription developed for GRBs in \citet{Granot2002_offaxis}, which assumes an emitting point source and accounts for the change in Doppler factor with viewing angle. Specifically, for viewing angle $\theta_\mathrm{obs}$ we can calculate the factor

\begin{equation}
    a(t) = \frac{1-\beta(t)}{1-\beta \cos{\theta_\mathrm{obs}}}.
\end{equation}

Then, the off-axis light curve is

\begin{equation}
    F_\mathrm{off} = a^3 F_{\mathrm{on},\nu/a} (at),
\end{equation}

\noindent where $F_\mathrm{on}$ in the expression above is calculated for a modified time $a(t)t$ and frequency $\nu_\mathrm{obs}/a(t)$. For a jet with identical properties to that of AT\,2022cmc, suppressing the optical emission by two orders of magnitude would require an off-axis parameter $a\approx5$. Off-axis TDE afterglow light curves have previously been explored in the context of Swift J1644+57 \citep{vanVelzen2013,Mimica2015,Generozov2017,Beniamini2023}. 








To explore the jet model in more detail, we calculate the optical ($g$ band; $6\times10^{14}\,$Hz) and radio (10\,GHz) light curves following the approach in \citet{Matsumoto2023} and \citet{Matsumoto2021}, which is based on standard gamma-ray burst afterglow theory \citep{Sari1998,Granot2002}. Interpreting the early optical emission as synchrotron emission from a jet afterglow, we have a spectral index that is quite steep. The best-fit spectral index to the LDT $gri$ epoch is $f_\nu \propto \nu^{-1.8\pm0.4}$, so electrons would likely be fast-cooling with $p\approx3$. As argued in \citet{Matsumoto2023}, the fact that the electrons are fast-cooling disfavors a reverse shock origin for the early optical flare, as this would predict a more abrupt decline in the early peak than what we observe. So, we assume that the early optical emission arises from a decelerating forward shock. 

In more detail, we assume a top-hat jet expanding into a medium with density profile

\begin{equation}
    n = n_0 (R/10^{17}\,\mathrm{cm})^{-k}.
\end{equation}

Assuming that the jet keeps its initial opening angle and neglecting the energy lost to radiation, energy conservation gives

\begin{equation}
E_{\rm k,iso}=\left[M_{\rm j,iso}+M_{\rm swept}(R)\Gamma]\right(\Gamma-1)c^2
\end{equation}

\noindent where 

\begin{equation}
M_{\rm swept}(R)\equiv\int_0^{R}4\pi m_p n(r)r^2dr
\end{equation}

\noindent and $M_{\rm j,iso}\equiv E_{\rm k,iso}/(\Gamma_0-1)c^2$ is the jet's isotropic-equivalent original mass. To an observer, the jet radius evolves according to 

\begin{equation}
\frac{dR}{dt}=\frac{\beta(R) c    }{1-\beta(R)}
\end{equation}

\noindent where time $t$ is the observer time and the denominator accounts for relativistic effects. We start the integration at a sufficiently short timescale that the jet may be regarded as freely expanding: $R_0=(\beta_0c)t_0/(1-\beta_0)$. 

Once we solve for the dynamics, i.e., the evolution of the blastwave Lorentz factor $\Gamma(t)$, we calculate the light curve as follows. The magnetic field is given by 

\begin{equation}
B=\sqrt{{32}\pi \epsilon_B m_p c^2 (\Gamma-1)\Gamma}
\end{equation}

\noindent where $\beta$ is the jet velocity. The characteristic and fast cooling electron Lorentz factors are given by 

\begin{equation}
\label{eq:gammam}
\gamma_{\rm m}={\rm max}\left[2,\frac{m_p}{4m_e}\bar{\epsilon_e} (\Gamma-1) \right]
\end{equation}

\noindent and

\begin{equation}
\gamma_{\rm c}=\frac{6\pi m_e c}{\sigma_{\rm T}B^2\delta_{\rm D} t},
\end{equation}

\noindent where $\bar{\epsilon_e}=4\epsilon_e \left(\frac{p-2}{p-1}\right)$ and $\delta_{\rm D}=\frac{1}{\Gamma(1-\beta)}$. For a given Lorentz factor $\gamma$ the corresponding synchrotron frequency is given by 

\begin{equation}
\nu_{\rm syn}(\gamma)=\delta_{\rm D}\frac{eB\gamma^2}{2\pi m_e c}\ .
\end{equation}

Since the synchrotron emissivity of a single electron with $\gamma_{\rm m}$ is given by

\begin{align}
P_{\nu_{\rm m}}=\delta_{\rm D}\Gamma\left[\frac{4}{3}\sigma_{\rm T}c\gamma_{\rm m}^2\frac{B^2}{8\pi}\right]\frac{1}{\nu_{\rm m}}\ ,
\end{align}

\noindent the flux at the frequency at which the bulk of electrons are emitting ($\nu_m$ in the slow cooling regime, $\nu_c$ in the fast cooling regime) is

\begin{equation}
F_{\nu_{\rm m}}=N_{\rm e}\frac{P_{\nu_{\rm m}}}{4\pi d_{\rm L}^2}{\rm min}[(\Gamma\theta_{\rm j})^2,1],
\end{equation}

\noindent and the number of electrons at that characteristic frequency is

\begin{equation}
N_{\rm e}=\frac{M_{\rm swept}(R)}{m_p}{\rm min}\left[1,\left(\frac{\beta}{\beta_{\rm DN}}\right)^2\right]
\end{equation}

\noindent where the last factor in $N_{\rm e}$ is needed to take into account for the deep Newtonian regime. Finally,

\begin{equation}
    \beta_{\mathrm{DN}} = \sqrt{\frac{16m_e}{\bar{\epsilon_e} m_p}}\ 
\end{equation}

\noindent is the critical velocity below which $\gamma_{\rm m}<2$ in Equation~\eqref{eq:gammam} and the deep-Newtonian regime sets in.

To calculate the spectrum neglecting synchrotron self-absorption, we follow the prescription of \cite{Granot2002}. More specifically, for the slow cooling regime of $\nu_{\rm m}<\nu_{\rm c}$ we use $F_\nu \propto \nu^{1/3}$ for $\nu<\nu_m$, $F_\nu \propto \nu^{-(p-1)/2}$ for $\nu_m < \nu < \nu_c$, and $F_\nu \propto \nu^{-p/2}$ for $\nu > \nu_c$. For the fast cooling regime of $\nu_{\rm m}>\nu_{\rm c}$,  we use $F_\nu \propto \nu^{1/3}$ for $\nu<\nu_c$, $F_\nu \propto \nu^{-1/2}$ for $\nu_c<\nu<\nu_m$, and $F_\nu \propto \nu^{-p/2}$ for $\nu>\nu_m$.
To connect the power-law segments, following \citet{Granot2002} we adopt a sharpness parameter $s$ (with the value obtained from their Table 2). 
In addition, to smoothly connect the fast and slow cooling regimes, we calculate $F_\nu$ for both cases and take a weighted average

\begin{equation}
F_\nu=\frac{\left(\frac{\nu_c}{\nu_m}\right)^2F_\nu^{\rm slow}+\left(\frac{\nu_m}{\nu_c}\right)^2F_\nu^{\rm fast}}{\left(\frac{\nu_c}{\nu_m}\right)^2+\left(\frac{\nu_m}{\nu_c}\right)^2}.
\end{equation}

\noindent Next, we account for synchrotron self-absorption. The absorption coefficient is \citep{Rybicki1986}

\begin{equation}
\alpha_\nu=\frac{\bar{p}\sqrt{3}e^3n{\rm min}\left[1,\left(\frac{\beta}{\beta_{\rm DN}}\right)^2\right]B\Gamma\delta_{\rm D}}{4\pi m_e^2c^2\gamma_{\rm m}\nu^2}
\left(\frac{\nu}{\nu_{\rm m}}\right)^{-p/2}I(\nu),
\end{equation}

\noindent where $\bar{p}=(p+2)(p-1)$ and $I(\nu)$ is the integral 

\begin{equation}
I(\nu)=\int_0^{\nu/\nu_{\rm m}}dx x^{\frac{p}{2}}\int_x^\infty dy K_{5/3}(y).
\end{equation}

\noindent Alternatively, in terms of the synchrotron function $F(x)$,

\begin{equation}
I(\nu)=\int_0^{\nu/\nu_{\rm m}}dx x^{\frac{p-2}{2}} F(x).
\end{equation}

We approximate $F(x)$ using the fitting function in \cite{Fouka2013}. For $\nu/\nu_m\geq10^{4}$, we use the following approximation for $I(\nu)$ from \citet{Rybicki1986}:

\begin{equation}
    I(\nu) = \frac{2^{\mu+1}}{\mu+2} 
    \Gamma \left( \frac{\mu}{2} + \frac{7}{3} \right) 
    \Gamma \left( \frac{\mu}{2} + \frac{2}{3} \right)
\end{equation}

\noindent where in this case $\mu=(p-2)/2$. We multiply the light curve by 

\begin{equation}
\frac{1-e^{-\tau}}{\tau}
\end{equation}

\noindent where $\tau = \alpha_{\nu}R$.

\begin{figure}
    \centering
    \includegraphics[width=\linewidth]{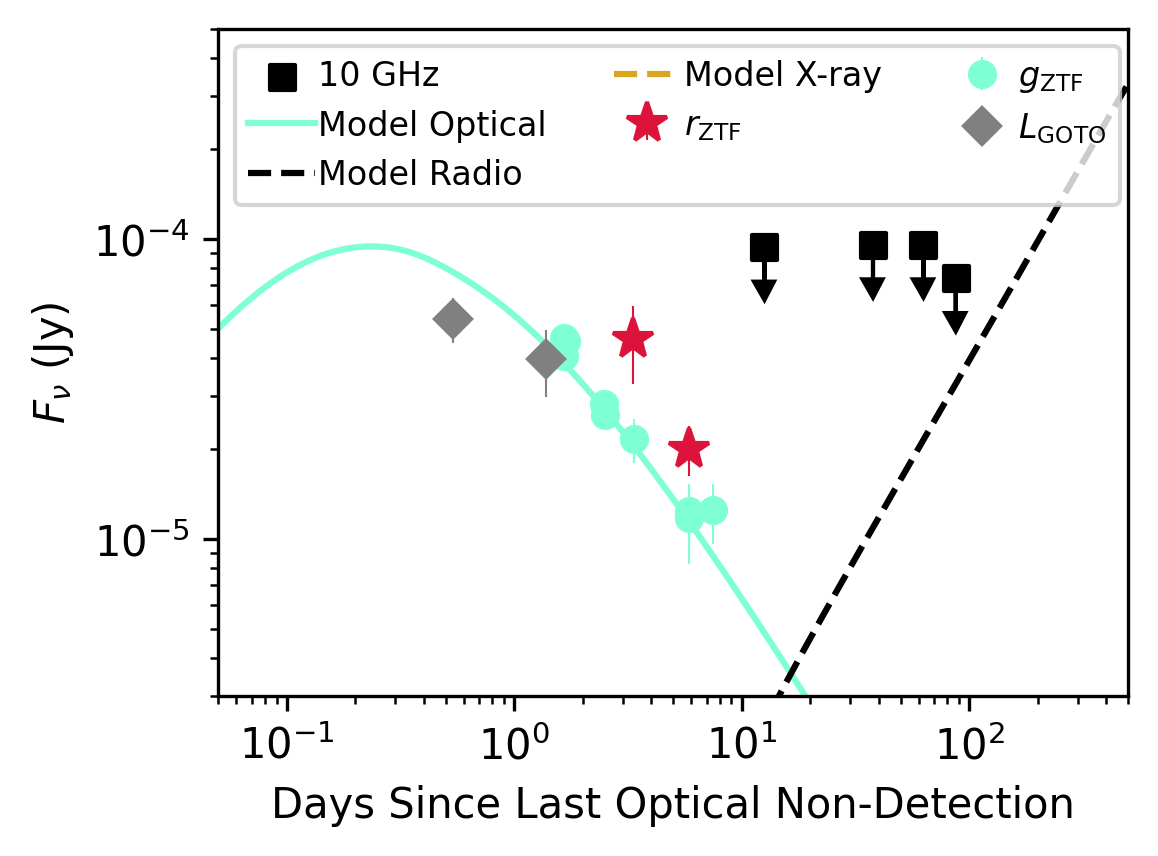}
    \caption{Model optical ($g$ band; solid line) and radio (10\,GHz; dashed line) light curves for an off-axis jet. Optical observations are plotted as circles ($g$ band), diamonds ($L$ band), and stars ($r$ band), while upper limits from 10\,GHz radio observations (assuming detections are from the host galaxy) are plotted as squares.}
    \label{fig:model}
\end{figure}

Figure~\ref{fig:model} shows a model with the following physical parameters: $\epsilon_B=0.002$, $\epsilon_e=0.2$, $E_{52}=5$, $\Gamma_0=3$, $n_{17}=20,000\,$cm$^{-3}$, $\theta_j=0.1$, $k=1$ where $n = n_{17} (r/10^{17}\,\mathrm{cm})^{-k}$, $p=2.2$, and $\theta_\mathrm{obs}=0.4$. 
The model underpredicts the observed X-ray luminosity by two orders of magnitude, consistent with previous work on jetted TDEs that attribute the hard variable X-rays to emission internal to the jet rather than to the forward shock \citep{Bloom2011,Zauderer2011,Crumley2016,Matsumoto2023,Yao2024}, or alternatively to an external reverse shock \citep{Yuan2024}. 

In this model, the early optical emission is suppressed by the fact that the jet is observed off-axis: the off-axis parameter $a$ varies from $a=0.5$ at $\Delta t=0.1\,$d to $a=0.9$ at $\Delta t=10\,$d. If this jet were viewed on-axis, the peak optical luminosity would be an order of magnitude fainter than that of AT\,2022cmc, primarily because the jet energy is an order of magnitude smaller. The viewing angle does not significantly suppress the radio luminosity; by our last radio observation, the shock is almost Newtonian. The radio peak is suppressed until late times by synchrotron self-absorption, so the timing of the radio peak is essentially set by the circumnuclear density. We note that a circumnuclear density of $20,000\,$cm$^{-3}$ at $10^{17}\,$cm (or 1000\,$R_s$ for a Schwarzschild radius of $R_s=10^{14}\,$cm, as expected for a $4\times10^{8}\,M_\odot$ black hole) is similar to that inferred for the TDE AT\,2019dsg \citep{Cendes2021_19dsg,Alexander2020}. Testing this model requires late-time radio observations to search for the predicted brightening in the radio light curve. 

\subsection{Shocks from Self-intersecting Debris}
\label{sec:model-streams}


The black hole in AT\,2024kmq is likely quite massive, $M\approx10^{8}\,M_\odot$ (Section~\ref{sec:host}). For the tidal disruption of a Sun-like star by such a massive black hole, the tidal radius $r_{\rm t} = R_{\star}\left(M/M_{\star}\right)^{1/3}$ is within a factor of a few of the gravitational radius $r_{\rm g} = GM/c^2$, meaning that general relativistic effects are important \citep{Hills1975, Stone2019}. 
Most relevant for this paper, for such highly relativistic tidal radii, apsidal precession amounts to many tens of degrees, which could drive debris stream collisions and thus energy loss and circularization \citep{Rees1988}---more massive black holes might therefore be expected to have more efficient disk formation. 

In more detail, 
the incoming and outgoing streams collide at large angles (i.e., not ``head on,'' as would be approximately the case for a lower-mass black hole). The incoming and outgoing material self-intersect over roughly the freefall time from the self-intersection radius, producing two shocks and a contact discontinuity separating the two. If the densities of the two streams are equal and the intersection angle is 90$^{\circ}$, symmetry demands that the contact discontinuity is angled at 45$^{\circ}$ with respect to the directions of the velocities of the streams at the point of contact. After approximately the freefall time from the self-intersection radius, the incoming debris stream must be replenished to initiate a second, strong self-intersection, which repeats until the disc is sufficiently dense that the incoming stream disintegrates via the Kelvin-Helmholtz instability and the system settles into more of a steady state \citep{andalman22}. 

The primary effect of the shocks will be to thermalize the gas to a pressure of $p \simeq \rho v^2$, where $\rho$ is the density and $v \simeq \mathrm{few}\times 0.1\, c$ is the velocity of the stream at the self-intersection point for these highly relativistic encounters. The shocked material then expands at approximately the original speed within the plane and at approximately the sound speed in the plane perpendicular to fill a quasi-spherical volume 
\citep{Jiang2016_self_crossing_shock, huang23, huang24}. \citet{Jiang2016_self_crossing_shock} and \citet{huang23} (see also \citealt{BonnerotLu2021} for analytical estimates) note that $\sim 10\%$ of the shocked gas can be ejected on positive-energy orbits, although the former study did not account for the gravitational field of the black hole, while the latter define unbound as being dictated by the positivity of an instantaneous Bernoulli-like parameter, and it is not clear that this parameter will be conserved given the ongoing time dependence (i.e., it is possible for the Bernoulli parameter to be positive but for material to remain bound; see \citealt{coughlin18} for a specific example). \citet{huang23} also note that their condition does not account for the directionality of the resulting flow, with some ``unbound'' debris directed toward the black hole. Nonetheless, it seems possible that some fraction of the debris is ejected on positive-energy orbits at speeds comparable to the speed at which gas self-intersects, which in the case of highly relativistic encounters is tens of percent the speed of light. Irrespective of its boundedness, a substantial amount of the debris recedes to distances comparable to the initial apocenter distance as set during the initial disruption process (i.e., the Keplerian energy), which is $R_{\star}\left(M/M_{\star}\right)^{2/3} \simeq 10^{16}M_{8}^{2/3}$ cm \citep{lacy82}.

The density of the stream as it self-intersects can be crudely approximated by treating it as a cylinder that has half the mass of the disrupted debris, a cross-sectional radius $R_{\star}$, and length equal to the apocenter of the most-bound debris, which yields $\rho \simeq 4\times 10^{-6} M_{8}^{-2/3}$ g cm$^{-3}$ for our set of parameters. However, this likely represents an over-estimate by at least 2--3 orders of magnitude, owing both to the fact that the stream extends well beyond the apocenter of the most-bound debris and because the less-dense tails return first. In any case, the shocked gas is highly optically thick and radiation-pressure dominated, and one can estimate the temperature from the shock jump conditions as $\rho v^2 \simeq aT^4/3$. For relativistic encounters the velocity at the self-intersection radius will be $\sim \mathrm{few}\times 0.1 c$ and independent of the black hole mass (i.e., the self-intersection occurs at only a few gravitational radii, so the velocity $v \sim \sqrt{GM/r}$ is independent of black hole mass), and letting $\rho = \rho_0 M_8^{-2/3}$ (where $\rho_0$ is likely between $10^{-8} - 10^{-10}$ g cm$^{-3}$, i.e., much less than the value of $10^{-6}$ g cm$^{-3}$ that comes from our simple geometrical estimate), we find $T = 1.4\times 10^{6} M_{8}^{-1/6}$ with $\rho_0 = 10^{-9}$ g cm$^{-3}$ and $v = 0.1 c$; this agrees with the temperatures inferred by \citet{Jiang2016_self_crossing_shock}. 

Following the self-intersection, shock heating redistributes the velocities and leads to a quasi-spherical expansion from the intersection point, as argued in the preceding paragraphs and as found in recent radiation-hydrodynamical simulations (in addition to those above, see \citealt{ramirez-ruiz09}). If the velocity profile of the expanding debris is approximately homologous and the density and pressure are $\sim$ uniform in the expanding debris, the temperature of the (radiation-pressure-dominated and optically thick) gas declines with the leading edge of the expanding cloud $R$ as $T \propto R^{-1}$. If the stream collision occurs at $5 GM/c^2$ and has an initial temperature of $10^6$ K, then by the time the cloud expands to the apocenter of the most-bound debris, the temperature decreases to $\sim 5\times 10^{3} M_8^{1/6}$ K. If the bulk of the emission continues to originate from near the edge of the cloud and the material is only marginally bound, we would expect\footnote{This temporal scaling assumes that the debris is still in the phase where it is moving out on a zero-energy orbit, i.e., $0.5(dR/dt)^2 \sim GM/R$; this is justified from the fact that $\mathrm{few}\times 10^{15}$ cm is still less than the apocenter distance of $\sim 10^{16}$ cm. If one instead assumes that the surface has reached the phase where it moves at a constant velocity, then the luminosity declines with time as $L\propto t^{-2}$.} the luminosity to decline with time as $L \propto R^2 T^4 \propto t^{-4/3}$, with a value of $L \simeq 9\times 10^{43} M_8^2$ erg s$^{-1}$ near the debris apocenter (this strong dependence on the mass does not extend to less relativistic encounters, where the self-intersection radius depends very strongly on the mass of the disrupting black hole).

This temperature near apocenter and the value and temporal decline of the luminosity are roughly consistent with the early and red outburst from AT\,2024kmq. While this temperature is lower than the values of $\sim \mathrm{few}\times 10^{4}$ K that are inferred from radiation-hydrodynamical simulations \citep{Jiang2016_self_crossing_shock}, the time taken for the debris to reach the apocenter distance of $\sim 10^{15}$ cm is $\lesssim 1$ day if it is moving at $0.3\,c$, and the earlier and hotter phases of the evolution could have been missed. 
If the early red outburst was due to the expanding cloud of debris produced from the initial stream self-intersection, it is natural to associate the second, larger and bluer outburst with fallback accretion onto the black hole. However, the time taken to reach the peak in the fallback rate following the disruption of a solar-mass star is \citep{bandopadhyay24} $\sim 20\times \left(M/10^6M_{\odot}\right)^{1/2} \simeq 200$ days for a $10^{8} M_{\odot}$ black hole, which is a factor of $\sim 5$ longer than the $\sim 40$ days observed (see Figure \ref{fig:opt_lc}). 
It has been argued that the rise in the fallback rate is steeper for partial TDEs \citep{nixon21}, and a partial TDE is more likely for such a massive black hole. However, the luminosity of this second blue peak is fairly typical for TDEs, so we have no evidence that the star was only partially stripped. 


Another challenge for the self-intersection model is that this mechanism is thought to be ubiquitous in TDEs, facilitating the circularization of the debris and subsequent accretion onto the black hole \citep{Rees1988}. So, a natural question to ask is why outbursts with similar properties are uncommon (Section~\ref{sec:rates}). One possibility is that most TDEs are powered by lower-mass black holes, for which the self-intersection occurs far less violently and at much larger (non-relativistic) radii\footnote{Another possibility is that the spin of the black hole and Lense-Thirring (nodal) precession results in no initial collision \citep{Guillochon2016}, with a self-intersection happening later and at a time when the density would have fallen by a much larger factor; qualitatively similar conclusions are reached when considering this possibility.}. Specifically, in the limit that the relativistic apsidal precession angle is small and given by the leading-order, relativistic value of $\delta\phi = 3GM/(r_{\rm t}c^2)$ (e.g., \citealt{carroll}), the self-intersection radius is $r_{\rm SI} \simeq 2r_{\rm t}/\delta \phi$, leading to a velocity at the point of self-intersection of $v_{\rm SI} \simeq \sqrt{GM/r_{\rm SI}} \simeq c\times\left(GM/(r_{\rm t}c^2)\right)$. The radiation pressure resulting from the shock is then $p_{\rm rad} \simeq a T^4 \simeq \rho c^2 \left(GM/(r_{\rm t}c^2)\right)^2$, which is reduced by a factor of $\left(GM/(r_{\rm t}c^2)\right)^2$ relative to the relativistic value discussed in the preceding paragraphs; for $M = 10^6M_{\odot}$ and a solar-like star, this factor is $\sim 4\times 10^{-4}$, such that the dissipation is far less efficient at powering a luminous outburst. It also seems likely that more material remains bound to the black hole (or at least ejected with a much smaller speed) following a comparatively less violent self-intersection, implying that most of the radiation is trapped within the flow that falls back to smaller radii where the process of accretion can liberate substantially more energy.


\section{Summary}
\label{sec:summary}

We presented the identification of AT\,2024kmq as a fast, red, and luminous optical transient, and its subsequent classification as a tidal disruption event by a massive (likely $M\approx10^8\,M_\odot$) black hole. With follow-up UV, optical, X-ray, and radio observations, we detected luminous and variable hard X-rays ($10^{44}\,$erg\,s$^{-1}$), featureless optical spectra, and set a limit on the presence of a radio counterpart ($L < 10^{39}\,$erg\,s$^{-1}$ at 10\,GHz from $\Delta t=10$--100\,d.) The radio emission detected at the transient position is likely from an underlying AGN; radio AGN are common in galaxies of this type. 

To our knowledge, an early optical peak with distinctly red colors in a TDE has only been seen in AT\,2022cmc, which is widely modeled as an on-axis jetted TDE. Using nearby TDEs classified as part of ZTF's flux-limited experiment, we find that early peaks with a similar luminosity and duration ($M<-19\,$mag for 1 week) are uncommon.

We consider two possibilities for the origin of the early red peak: an off-axis jet, and self-intersecting debris streams leading to the formation of the accretion disk. The off-axis jet model is motivated by AT\,2022cmc and the suggestion in the literature that ``featureless TDEs'' are associated with jets. For a jet, we find that an off-axis viewing angle ($\theta_\mathrm{obs}=0.4$, which is $4\times$ the jet opening angle) and a low jet energy ($E_j=5\times10^{52}\,$erg, an order of magnitude lower than that of AT\,2022cmc) are required to suppress the luminosity of the optical peak relative to what was observed in AT\,2022cmc. In addition, a circumnuclear density of $n_{17}=20,000\,$cm$^{-3}$ is required to suppress the radio emission until late times. Under this model, the radio emission should eventually be detectable, on timescales of years. 

For the self-intersecting debris stream model, we find that the highly relativistic nature of the encounter (due to the high mass of the black hole) could produce optical emission with a similar temperature and luminosity. In this model, it is natural to explain the second blue peak as fallback accretion onto the black hole, although the timescale is challenging to explain. Another challenge for this model is to explain why such luminous optical flares are relatively uncommon; one possible explanation is that they are much fainter in encounters involving lower-mass black holes. 

\vspace{1cm}
\noindent \emph{Acknowledgements.}

AYQH would like to thank Eliot Quataert, Dong Lai, and Wenbin Lu for helpful conversations regarding TDE models; as well as Krista Lynne Smith for discussions about radio AGN. ERC acknowledges support from NASA through the Astrophysics Theory Program, grant 80NSSC24K0897. JDL acknowledges support from a UK Research and Innovation Future Leaders Fellowship (MR/T020784/1). I.A. is supported by NSF AST2407924 and NASA ADAP24-0159.
C.L. is supported by DoE award \#DE-SC0025599. J.S. is supported by NASA award 80NSSC24K0377.

This work was performed in part at the Aspen Center for Physics, which is supported by National Science Foundation grant PHY-2210452.

Based on observations obtained with the Samuel Oschin Telescope 48-inch and the 60-inch Telescope at the Palomar Observatory as part of the Zwicky Transient Facility project. ZTF is supported by the National Science Foundation under Grant No. AST-2034437 and a collaboration including Caltech, IPAC, the Oskar Klein Center at Stockholm University, the University of Maryland, University of California, Berkeley , the University of Wisconsin at Milwaukee, University of Warwick, Ruhr University Bochum, Cornell University, Northwestern University and Drexel University. Operations are conducted by COO, IPAC, and UW.
The ZTF forced-photometry service was funded under the Heising-Simons Foundation grant No. 12540303 (PI: Graham).
The Gordon and Betty Moore Foundation, through both the Data-Driven Investigator Program and a dedicated grant, provided critical funding for SkyPortal.


SED Machine is based upon work supported by the National Science Foundation under Grant No. 1106171. 
Data were obtained at the Lick Observatory, which is a multi-campus research unit of the University of California.
Some of the data presented herein were obtained at the W. M. Keck Observatory, which is operated as a scientific partnership among the California Institute of Technology, the University of California, and NASA. The Observatory was made possible by the generous financial support of the W. M. Keck Foundation.
The authors wish to recognise and acknowledge the very significant cultural role and reverence that the summit of Maunakea has always had within the indigenous Hawaiian community. We are most fortunate to have the opportunity to conduct observations from this mountain.

Based on observations carried out with the IRAM Interferometer NOEMA. IRAM is supported by INSU/CNRS (France), MPG (Germany) and IGN (Spain).

This work made use of data supplied by the UK Swift Science Data Centre at the University of Leicester.

This work uses data obtained with eROSITA telescope onboard SRG observatory. The SRG observatory was built by Roskosmos with the participation of the Deutsches Zentrum für Luft- und Raumfahrt (DLR). The SRG/eROSITA X-ray telescope was built by a consortium of German Institutes led by MPE, and supported by DLR. The SRG spacecraft was designed, built, launched and is operated by the Lavochkin Association and its subcontractors. The science data were downlinked via the Deep Space Network Antennae in Bear Lakes, Ussurijsk, and Baykonur, funded by Roskosmos. The eROSITA data used in this work were processed using the eSASS software system developed by the German eROSITA consortium and proprietary data reduction and analysis software developed by the Russian eROSITA Consortium.


The National Radio Astronomy Observatory is a facility of the National Science Foundation operated under cooperative agreement by Associated Universities, Inc.

LOFAR is the Low Frequency Array designed and constructed by ASTRON. It has observing, data processing, and data storage facilities in several countries, which are owned by various parties (each with their own funding sources), and which are collectively operated by the ILT foundation under a joint scientific policy. The ILT resources have benefited from the following recent major funding sources: CNRS-INSU, Observatoire de Paris and Université d'Orléans, France; BMBF, MIWF-NRW, MPG, Germany; Science Foundation Ireland (SFI), Department of Business, Enterprise and Innovation (DBEI), Ireland; NWO, The Netherlands; The Science and Technology Facilities Council, UK; Ministry of Science and Higher Education, Poland; The Istituto Nazionale di Astrofisica (INAF), Italy.

This research made use of the Dutch national e-infrastructure with support of the SURF Cooperative (e-infra 180169) and the LOFAR e-infra group. The Jülich LOFAR Long Term Archive and the German LOFAR network are both coordinated and operated by the Jülich Supercomputing Centre (JSC), and computing resources on the supercomputer JUWELS at JSC were provided by the Gauss Centre for Supercomputing e.V. (grant CHTB00) through the John von Neumann Institute for Computing (NIC).

This research made use of the University of Hertfordshire high-performance computing facility and the LOFAR-UK computing facility located at the University of Hertfordshire and supported by STFC [ST/P000096/1], and of the Italian LOFAR IT computing infrastructure supported and operated by INAF, and by the Physics Department of Turin university (under an agreement with Consorzio Interuniversitario per la Fisica Spaziale) at the C3S Supercomputing Centre, Italy.

This scientific work uses data obtained from Inyarrimanha Ilgari Bundara, the CSIRO Murchison Radio-astronomy Observatory. We acknowledge the Wajarri Yamaji People as the Traditional Owners and native title holders of the Observatory site. CSIRO’s ASKAP radio telescope is part of the Australia Telescope National Facility (https://ror.org/05qajvd42). Operation of ASKAP is funded by the Australian Government with support from the National Collaborative Research Infrastructure Strategy. ASKAP uses the resources of the Pawsey Supercomputing Research Centre. Establishment of ASKAP, Inyarrimanha Ilgari Bundara, the CSIRO Murchison Radio-astronomy Observatory and the Pawsey Supercomputing Research Centre are initiatives of the Australian Government, with support from the Government of Western Australia and the Science and Industry Endowment Fund.

This paper includes archived data obtained through the CSIRO ASKAP Science Data Archive, CASDA (http://data.csiro.au).


\software{
CASA \citep{McMullin2007},
Astropy \citep{astropy:2013, astropy:2018, astropy:2022}
}

\facilities{
PO:1.2m, 
Keck:I,
Shane,
Swift,
EVLA,
LDT,
Liverpool:2m
}

\clearpage
\newpage
\appendix

\section{Additional Data Tables}
\label{sec:appendix}

We present the optical photometry in Table~\ref{tab:optical-photometry} and the radio observations in Table~\ref{tab:radio}.

\startlongtable
\begin{deluxetable}{llllll}
\tablecolumns{6}
\tablecaption{Optical photometry of AT\,2024kmq. \label{tab:optical-photometry}}
\tablehead{
\colhead{Start MJD} & 
\colhead{$\Delta t_\mathrm{obs}$ (d)} & \colhead{Instrument} & 
\colhead{Filter} & 
\colhead{Mag$^{a}$} & \colhead{eMag}}
\startdata
60460.88621 & 0.64 & GOTO & $L$ & 19.63 & 0.19 \\
60461.88658 & 1.64 & GOTO & $L$ & 19.97 & 0.27 \\
60462.22313 & 1.98 & ZTF/P48$^b$ & $g$ & 19.95 & 0.08 \\
60462.22406 & 1.98 & ZTF/P48 & $g$ & 19.80 & 0.08 \\
60462.25939 & 2.01 & ZTF/P48 & $g$ & 19.83 & 0.07 \\
60463.19822 & 2.95 & ZTF/P48 & $g$ & 20.35 & 0.12 \\
60463.21853 & 2.97 & ZTF/P48 & $g$ & 20.44 & 0.11 \\
60464.21958 & 3.97 & ZTF/P48 & $r$ & 19.79 & 0.32 \\
60464.23774 & 3.99 & ZTF/P48 & $g$ & 20.65 & 0.18 \\
60467.21535 & 6.97 & LDT/LMI & $r$ & 20.70 & 0.04 \\
60467.21888 & 6.97 & P60/SEDM & $g$ & 21.25 & 0.13 \\
60467.22667 & 6.98 & LDT/LMI & $i$ & 20.50 & 0.07 \\
60467.23357 & 6.99 & LDT/LMI & $g$ & 21.30 & 0.12 \\
60469.18617 & 8.94 & ZTF/P48 & $g$ & 21.24 & 0.25 \\
60479.21865 & 18.97 & ZTF/P48 & $g$ & 20.61 & 0.32 \\
60480.20184 & 19.96 & ZTF/P48 & $g$ & 20.50 & 0.30 \\
60480.23791 & 19.99 & ZTF/P48 & $g$ & 20.45 & 0.34 \\
60481.19624 & 20.95 & ZTF/P48 & $g$ & 20.45 & 0.26 \\
60481.26595 & 21.02 & ZTF/P48 & $g$ & 20.19 & 0.36 \\
60482.21664 & 21.97 & ZTF/P48 & $g$ & 20.31 & 0.30 \\
60483.19494 & 22.95 & ZTF/P48 & $g$ & 19.81 & 0.19 \\
60483.24196 & 23.00 & ZTF/P48 & $g$ & 20.07 & 0.29 \\
60484.26888 & 24.02 & ZTF/P48 & $g$ & 19.41 & 0.19 \\
60486.20213 & 25.96 & ZTF/P48 & $g$ & 19.44 & 0.06 \\
60489.22517 & 28.98 & ZTF/P48 & $g$ & 18.88 & 0.05 \\
60490.96002 & 30.72 & GOTO & $L$ & 19.07 & 0.12 \\
60492.91381 & 32.67 & GOTO & $L$ & 18.79 & 0.09 \\
60495.89886 & 35.65 & GOTO & $L$ & 18.58 & 0.10 \\
60499.19725 & 38.95 & ZTF/P48 & $g$ & 18.11 & 0.03 \\
60499.89569 & 39.65 & GOTO & $L$ & 18.15 & 0.08 \\
60500.20609 & 39.96 & ZTF/P48 & $g$ & 18.10 & 0.03 \\
60500.22086 & 39.98 & ZTF/P48 & $g$ & 18.08 & 0.04 \\
60501.20279 & 40.96 & ZTF/P48 & $g$ & 17.98 & 0.03 \\
60501.21436 & 40.97 & ZTF/P48 & $g$ & 17.98 & 0.03 \\
60501.89200 & 41.65 & GOTO & $L$ & 18.40 & 0.09 \\
60501.93506 & 41.69 & LT/IO:O & $g$ & 18.15 & 0.17 \\
60501.93629 & 41.69 & LT/IO:O & $r$ & 18.29 & 0.09 \\
60501.93749 & 41.69 & LT/IO:O & $i$ & 18.74 & 0.09 \\
60501.93870 & 41.69 & LT/IO:O & $z$ & 18.17 & 0.13 \\
60501.93996 & 41.70 & LT/IO:O & $u$ & 17.80 & 0.22 \\
60502.20354 & 41.96 & ZTF/P48 & $g$ & 18.06 & 0.04 \\
60502.23672 & 41.99 & P60/SEDM & $g$ & 18.05 & 0.06 \\
60502.23940 & 41.99 & P60/SEDM & $r$ & 18.29 & 0.06 \\
60502.24210 & 42.00 & P60/SEDM & $i$ & 18.42 & 0.03 \\
60502.95101 & 42.71 & LT/IO:O & $r$ & 18.32 & 0.07 \\
60502.95316 & 42.71 & LT/IO:O & $g$ & 18.13 & 0.18 \\
60502.95509 & 42.71 & LT/IO:O & $i$ & 18.51 & 0.09 \\
60502.95722 & 42.71 & LT/IO:O & $z$ & 18.75 & 0.15 \\
60502.95941 & 42.71 & LT/IO:O & $u$ & 17.99 & 0.20 \\
60503.18232 & 42.94 & P60/SEDM & $g$ & 18.15 & 0.06 \\
60503.18500 & 42.94 & P60/SEDM & $r$ & 18.21 & 0.04 \\
60503.18559 & 42.94 & ZTF/P48 & $g$ & 18.06 & 0.04 \\
60503.18770 & 42.94 & P60/SEDM & $i$ & 18.54 & 0.04 \\
60503.20007 & 42.96 & ZTF/P48 & $g$ & 17.99 & 0.05 \\
60505.19919 & 44.95 & ZTF/P48 & $g$ & 17.96 & 0.04 \\
60505.21012 & 44.97 & P60/SEDM & $g$ & 18.08 & 0.04 \\
60505.21281 & 44.97 & P60/SEDM & $r$ & 18.19 & 0.04 \\
60505.21550 & 44.97 & P60/SEDM & $i$ & 18.44 & 0.06 \\
60506.90029 & 46.66 & GOTO & $L$ & 18.27 & 0.22 \\
60509.20082 & 48.96 & ZTF/P48 & $g$ & 18.08 & 0.06 \\
60510.19566 & 49.95 & ZTF/P48 & $g$ & 18.09 & 0.06 \\
60510.52183 & 50.28 & Swift/UVOT & $UVM2$ & 18.15 & 0.05 \\
60511.90682 & 51.66 & LT/IO:O & $g$ & 18.19 & 0.10 \\
60511.90783 & 51.66 & LT/IO:O & $i$ & 18.72 & 0.08 \\
60511.90885 & 51.66 & LT/IO:O & $r$ & 18.30 & 0.12 \\
60511.90984 & 51.66 & LT/IO:O & $z$ & 18.38 & 0.15 \\
60512.90387 & 52.66 & Swift/UVOT & $UVW1$ & 17.76 & 0.06 \\
60512.90530 & 52.66 & Swift/UVOT & $U$ & 17.73 & 0.07 \\
60512.90628 & 52.66 & Swift/UVOT & $B$ & 17.96 & 0.14 \\
60512.90866 & 52.66 & Swift/UVOT & $UVW2$ & 18.01 & 0.05 \\
60512.91102 & 52.67 & Swift/UVOT & $V$ & 18.33 & 0.33 \\
60512.91262 & 52.67 & Swift/UVOT & $UVM2$ & 17.88 & 0.06 \\
60514.65172 & 54.41 & Swift/UVOT & $UVW1$ & 17.73 & 0.07 \\
60514.65262 & 54.41 & Swift/UVOT & $U$ & 17.62 & 0.08 \\
60514.65369 & 54.41 & Swift/UVOT & $UVW2$ & 18.07 & 0.06 \\
60514.68398 & 54.44 & Swift/UVOT & $UVM2$ & 17.92 & 0.06 \\
60518.92244 & 58.68 & Swift/UVOT & $UVM2$ & 18.10 & 0.08 \\
60518.92483 & 58.68 & Swift/UVOT & $UVW1$ & 17.88 & 0.08 \\
60518.92629 & 58.68 & Swift/UVOT & $U$ & 17.86 & 0.10 \\
60518.92835 & 58.68 & Swift/UVOT & $UVW2$ & 18.26 & 0.07 \\
60519.18373 & 58.94 & ZTF/P48 & $g$ & 18.17 & 0.04 \\
60519.19725 & 58.95 & ZTF/P48 & $g$ & 18.35 & 0.07 \\
60519.90265 & 59.66 & LT/IO:O & $g$ & 18.24 & 0.10 \\
60519.90493 & 59.66 & LT/IO:O & $i$ & 18.96 & 0.09 \\
60519.90734 & 59.66 & LT/IO:O & $r$ & 18.45 & 0.04 \\
60519.90961 & 59.66 & LT/IO:O & $z$ & 18.71 & 0.14 \\
60520.23891 & 59.99 & Swift/UVOT & $UVM2$ & 18.15 & 0.07 \\
60520.24083 & 60.00 & Swift/UVOT & $UVW1$ & 17.95 & 0.07 \\
60520.24202 & 60.00 & Swift/UVOT & $U$ & 17.91 & 0.08 \\
60520.24340 & 60.00 & Swift/UVOT & $UVW2$ & 18.28 & 0.06 \\
60521.17615 & 60.93 & P60/SEDM & $r$ & 18.40 & 0.05 \\
60521.17885 & 60.93 & P60/SEDM & $i$ & 18.45 & 0.05 \\
60521.89335 & 61.65 & LT/IO:O & $g$ & 18.33 & 0.12 \\
60521.89629 & 61.65 & LT/IO:O & $r$ & 18.42 & 0.03 \\
60521.89739 & 61.65 & LT/IO:O & $i$ & 18.71 & 0.10 \\
60521.89952 & 61.65 & LT/IO:O & $z$ & 18.85 & 0.10 \\
60528.88471 & 68.64 & LT/IO:O & $g$ & 18.60 & 0.10 \\
60528.88572 & 68.64 & LT/IO:O & $r$ & 18.53 & 0.05 \\
60528.88784 & 68.64 & LT/IO:O & $i$ & 19.01 & 0.07 \\
60528.88998 & 68.65 & LT/IO:O & $z$ & 19.24 & 0.11 \\
\enddata
\tablenotetext{a}{Not corrected for Milky Way extinction}
\tablenotetext{b}{From forced photometry on ZTF images}
\end{deluxetable}

\startlongtable
\begin{deluxetable*}{llllll}
\tablecolumns{6}
\tablecaption{Very Large Array observations of AT\,2024kmq. 
\label{tab:radio}}
\tablehead{\colhead{Start MJD} & \colhead{$\Delta t_\mathrm{obs}$} & \colhead{$\nu_\mathrm{obs}$} & \colhead{Beam Size} & \colhead{Beam Angle} & \colhead{$f_\nu$ (point source)$^a$} 
\\ 
\colhead{} & \colhead{(d)} & \colhead{(GHz)} & \colhead{(arcsec)} & \colhead{(deg)} & \colhead{(mJy)} 
}
\startdata
60475.98 & 15.09 & 10 & $0.66 \times 0.56$ & -64.31 & $0.112 \pm 0.01$ \\
60505.82 & 44.93 & 6 & $1.34 \times 1.06$ & 81.57 & $0.191 \pm 0.011$ \\
	 & 	 & 10 & $0.73 \times 0.56$ & 88.88 & $0.114 \pm 0.012$ \\
	 & 	 & 15 & $0.54 \times 0.39$ & -71.89 & $0.095 \pm 0.011$ \\
60536.08 & 75.19 & 6 & $1.55 \times 0.94$ & -75.48 & $0.172 \pm 0.014$ \\
	 & 	 & 10 & $0.93 \times 0.57$ & -75.53 & $0.114 \pm 0.013$ \\
60565.03 & 104.14 & 6 & $1.93 \times 0.93$ & -70.36 & $0.152 \pm 0.017$ \\
	 & 	 & 10 & $1.16 \times 0.62$ & -66.09 & $0.088 \pm 0.013$ \\
	 & 	 & 15 & $0.91 \times 0.39$ & -68.0 & $0.048 \pm 0.01$ \\
\enddata
\tablenotetext{a}{$f_\nu$ is measured using \texttt{imtool}, which allows for a forced point source fitting. When left to freely fit the shape, there are a few epochs where \texttt{imtool} prefers an extended shape, however for consistency we only quote the measured $f_\nu$ from the forced point source fitting, which is not significantly different from the measured $f_\nu$ when an extended shape is preferred.}
\tablenotetext{b}{The emission likely arises from the host galaxy, given that the flux density values do not change significantly over the course of our observations.}
\end{deluxetable*}

\bibliography{tde}{}
\bibliographystyle{aasjournal}

\end{document}